\documentclass[10pt]{article}

\usepackage{graphicx} % Required for inserting images
\usepackage{amsmath}
\usepackage{lipsum}
\usepackage{pdfcomment}

\usepackage{hyperref} % Include the hyperref package
\usepackage[english]{babel}
\usepackage[utf8]{inputenc}
\usepackage{amsfonts}
\usepackage{longtable}
\usepackage{booktabs}
\usepackage{float} 
\usepackage{geometry}
\usepackage{lineno}
\usepackage{authblk}

\usepackage{mathpazo}

\usepackage{array}
\usepackage{multirow}

\usepackage{amssymb}
\usepackage{pifont}

\usepackage[colorinlistoftodos]{todonotes}

\usepackage{wrapfig}

\usepackage[numbers,sort&compress]{natbib}
\usepackage{cite}

\hypersetup{
    colorlinks=true, % Enable colored links
    linkcolor=black,  % Set the color of internal links (e.g., references)
    citecolor=blue, % Change this color as desired
    urlcolor=blue    % Set the color of external links (e.g., URLs)
}

\usepackage{amsthm}
\usepackage{titlesec}
\usepackage{titletoc}
\titleclass{\subsubsubsection}{straight}[\subsection]
\newcounter{subsubsubsection}[subsubsection]
\renewcommand\thesubsubsubsection{\thesubsubsection.\arabic{subsubsubsection}}
\titleformat{\subsubsubsection}{\normalfont\normalsize\bfseries}{\thesubsubsubsection}{1em}{}
\titlespacing*{\subsubsubsection}{0pt}{3.25ex plus 1ex minus .2ex}{1.5ex plus .2ex}
\titleformat*{\section}{\fontsize{11}{12}\selectfont\bfseries}
\titleformat*{\subsection}{\fontsize{11}{12}\selectfont\bfseries}
\titleformat*{\subsubsection}{\fontsize{11}{12}\selectfont\bfseries}
\titleformat*{\paragraph}{\fontsize{11}{12}\selectfont\bfseries}
\titleformat*{\subparagraph}{\fontsize{11}{12}\selectfont\bfseries}

\titlecontents{subsubsubsection}
[8em] % adjust this value to fit your needs
{\small}
{\hyperlink{toc.\thecontentslabel}{\thecontentslabel.} }
{}
{\ \titlerule*[.5pc]{.}\contentspage}

\titlespacing*{\subsubsubsection}{0pt}{3.25ex plus 1ex minus .2ex}{1.5ex plus .2ex}

\usepackage{caption}
\usepackage{microtype}

\usepackage{enumitem}
\setlist{itemsep=0em}

\usepackage{fancyhdr} % For custom headers/footers
\usepackage{lastpage} % For total number of pages

\usepackage{listings}
\usepackage{xcolor}

\definecolor{codegreen}{rgb}{0,0.6,0}
\definecolor{codegray}{rgb}{0.5,0.5,0.5}
\definecolor{codepurple}{rgb}{0.58,0,0.82}
\definecolor{backcolour}{rgb}{0.95,0.95,0.92}

\lstdefinestyle{mystyle}{
    backgroundcolor=\color{backcolour},   
    commentstyle=\color{codegreen},
    keywordstyle=\color{magenta},
    numberstyle=\tiny\color{codegray},
    stringstyle=\color{codepurple},
    basicstyle=\ttfamily\footnotesize,
    breakatwhitespace=false,         
    breaklines=true,                 
    captionpos=b,                    
    keepspaces=true,                 
    numbers=left,                    
    numbersep=5pt,                  
    showspaces=false,                
    showstringspaces=false,
    showtabs=false,                  
    tabsize=2
}

\title{Assessing the Influence of Pavement Performance on Road Safety Through Crash Frequency and Severity Analysis}
\date{}

\author[1]{\normalsize Prathyush Kumar Reddy Lebaku}
\author[1]{Lu Gao \thanks{lgao5@central.uh.edu}}
\author[2]{Jingran Sun}
\author[3]{Xingju Wang}
\author[3]{Xuejian Kang}
\affil[1]{Department of Civil and Environmental Engineering, University of Houston}
\affil[2]{Center for Transportation Research, The University of Texas at Austin}
\affil[3]{School of Traffic and Transportation, Shijiazhuang Tiedao University}

\begin{document}

% \linenumbers

\maketitle

\abstract{Road safety is impacted by a range of factors that can be categorized into human, vehicle, and roadway/environmental elements. This research explores the connection between pavement performance and road safety, particularly in relation to crash frequency and severity, using data from the Iowa Department of Transportation (DOT) for 2022. By merging crash data with pavement inventory data, we conduct a spatial analysis that incorporates the geographical coordinates of crash sites with the conditions of road segments. Statistical methods are applied to compare crash rates and severity across various pavement condition categories. To identify the most influential factors affecting crash rates and severity, we use machine learning models along with negative binomial and ordered probit regression models. The study's key findings reveal that higher speed limits, well-maintained roads, and improved friction scores correlate with lower crash rates, whereas rougher roads and adverse weather conditions are linked to higher crash severity. This analysis emphasizes the critical need for prioritizing pavement maintenance and integrating safety-focused design principles to boost road safety. Moreover, the study underscores the ongoing need for research to better understand and address the intricate relationship between pavement performance and road safety.}



% \linenumbers
\maketitle

\section{Introduction}
Road safety is influenced by a myriad of factors that can be broadly categorized into three groups: human factors, vehicle factors, and roadway/environmental factors \citep{megnidio2023machine,wegman2017future,hughes2015review,smeed1949some,toroyan2009global,world2020road,ziakopoulos2020review}. Previous studies have demonstrated that deteriorated pavement quality contributes to hazardous driving conditions, increasing the likelihood of accidents \citep{li2013impact,lee2015effects, chan2009relationship, buddhavarapu2013influence, li2014safety, chan2010investigating, mayora2009assessment, zeng2014estimation, kopelias2007urban}. For instance, \citet{elghriany2016investigation} and \citet{chen2018comparative} found a correlation between pavement roughness and crash rates, with higher roughness leading to increased crash frequency. Similarly, \citet{ali2023evaluating} found that defects such as potholes, cracks, and uneven surfaces pose substantial risks to motorists, potentially causing vehicle control loss or collisions. To address such hazards, systematic pavement management programs that continuously monitor pavement conditions, and that prioritize maintenance and rehabilitation activities, are essential for reducing crash likelihood and improving overall road safety  \citep{lebaku2024deep,tighe2000incorporating, gao2012network,gao2008robust,pasindu2022incorporating,gao2022missing,haas1978pavement,gao2022evaluating}. In line with this, \citet{rafiei2024effect, jia2022evaluation} investigated the impact of rutting depth and width to the roadway safety and driving quality of vehicles. Moreover, pavement skid resistance and macrotexture have been found to be related to crash risk \citep{gao2015milled}. In addition to pavement condition, road geometric factors such as alignment, curvature, and slope significantly influence driver behavior and maneuverability. Researches indicate that road geometry directly affects safety outcomes, with sharp curves, inadequate visibility, and steep slopes increasing accident risk \citep{karlaftis2002effects,cafiso2021crash}. These factors become even more critical during adverse weather conditions or night-time driving \citep{becker2022weather,jagerbrand2016effects}. And roadway geometric data has also been combined with friction data to analyze their impacts on traffic crash \citep{roy2023effects}. Integrating road geometric factors into safety analyses provides a holistic understanding of the complex dynamics between pavement performance and road safety. Moreover, research indicates that increasing speed limits can lead to higher crash severity and increased fatalities. For instance, it was found that there is a 35.8\% rise in fatal crashes when speed limits were increased from 55 to 65 mph, and a 33.9\% rise when increased from 65 to 70 mph \citep{musabbir2020assessing}. In Texas, researchers observed a 39.2\% increase in crashes involving overturned vehicles and a 10.6\% rise in property damage-only crashes after raising speed limits to 75 mph on rural freeways \citep{avelar2024assessing}.

In examining the relationship between pavement performance and safety, many studies conducted in the past focused primarily targeting two variables: crash frequency and crash severity. Here, we provide a summary of key findings related to the impact of pavement condition on road safety, incorporating various factors, including pavement condition and road geometric factors, as reported in previous studies.

\subsection{Crash Rates}
The impact of the International Roughness Index (IRI) on crash rates varies among studies. Overall, higher IRI values tend to be associated with increased accident rates, particularly for multi-vehicle accidents, with some studies suggesting the existence of a threshold value below which crash rates do not significantly increase. \citet{al-masaeid1997ImpactPavementCondition} found that an increase in IRI reduces the single-vehicle accident rate but increases the multi-vehicle accident rate. \citet{chan2010InvestigatingEffectsAsphalt} and \citet{anastasopoulos2012StudyFactorsAffecting} also reported that higher IRI values increase all types of accident rates. \citet{li2014SafetyImpactPavement} demonstrated that low ride scores are associated with higher crash rates than fair and good ride scores.  \citet{vinayakamurthy2017EffectPavementCondition} identified a threshold IRI value of 210 in/mile, above which crash rates start to increase. \citet{ghanbari2017ImpactPavementCondition} found that the number of collisions tends to increase with rougher roads, with the impact of road roughness on multi-vehicle accidents being twice that on single-vehicle accidents. \citet{li2022statistical} confirmed that IRI positively impacts crash frequency.

Rutting depth has consistently been associated with heightened crash rates, particularly in adverse conditions such as nighttime and rainfall. However, research suggests that these rates may not surge until rutting exceeds a specific threshold, typically around 0.4 inches. \citet{graves2005MiningAnalysisTraffic} demonstrated that an increase in rutting depth corresponds to increased crash rates. \citet{chan2010InvestigatingEffectsAsphalt} noted that elevated rutting depth amplifies crash rates, especially during nighttime and rainy conditions. \citet{anastasopoulos2012StudyFactorsAffecting} found that escalating rutting depth correlates with heightened accident rates. \citet{vinayakamurthy2017EffectPavementCondition} highlighted a critical threshold effect, indicating that crash rates remain relatively stable until rut depth surpasses approximately 0.4 inches, beyond which they begin to rise. However, \citet{ghanbari2017ImpactPavementCondition} discovered that rut depth was not a significant factor in collision occurrences on rural roads, suggesting potential variations in the impact of rutting depth across different road types and settings.

For general pavement condition index, roads exhibiting very poor conditions shows significantly elevated crash rates compared to those in good condition, with the former experiencing crash rates more than double that of the latter, as noted by \citet{li2014SafetyImpactPavement}. They also observed that roadways with poor to fair condition scores displayed comparable crash rates. However, the impact of poor road conditions is more pronounced in certain types of crashes, particularly evident on medium-speed urban and rural collector roads. The research indicates that poor condition scores disproportionately affect specific categories of crashes. The authors found that these conditions impact same-direction crashes, incidents with  no injuries, crashes occurring in good pavement and light conditions, and those on rural and urban collectors of medium speed limits. \citet{adeli2022influence} showed that fractured particles and coarse aggregates have a decreasing effect, while binder content has an increasing effect on accident rates.

Pavement width appears to have no effect on accident rates, as found by \citet{al-masaeid1997ImpactPavementCondition}. The width of road shoulders also has no noticeable impact on accident rates. However, the type of shoulder surface material influences accident occurrences, with grass-type shoulders associated with higher incident rates, as noted by \citet{graves2005MiningAnalysisTraffic}. Despite this, the overall presence of a shoulder tends to decrease crash rates, with \citet{li2022statistical} noting a negative coefficient on crash rates in the presence of shoulders.

The occurrence of vertical curves on roads has been associated with increased multi-vehicle crash rates, as highlighted by \citet{al-masaeid1997ImpactPavementCondition}. Additionally, a steeper average gradient or grade has been found to contribute to higher crash rates, as shown by \citet{ghanbari2017ImpactPavementCondition, li2022statistical}. In terms of horizontal curves, their impact on accident rates varies. \citet{al-masaeid1997ImpactPavementCondition} noted that a higher number of curves tends to result in more single-vehicle accidents. Conversely, \citet{anastasopoulos2012StudyFactorsAffecting} discovered that the presence of horizontal curves might reduce crash rates. However, \citet{li2022statistical} demonstrated that longer curves tend to be associated with higher crash rates.

When it comes to skid resistance, research by \citet{li2022statistical} indicates that crashes decrease gradually, ranging from 2.5\% to 0.9\% per unit of friction in skid number. Regarding speed limits, their influence on accident rates varies according to different studies. \citet{al-masaeid1997ImpactPavementCondition} suggests that speed limits do not significantly affect accident rates. Conversely, \citet{ghanbari2017ImpactPavementCondition} discovered that for multi-vehicle crashes, higher average posted speed limits have a negative impact. These differing conclusions highlight the intricate nature of the relationship between speed limits and accident rates, suggesting the necessity for further investigation to road safety management. As for the percentage of trucks, \citet{al-masaeid1997ImpactPavementCondition} found that it has no noticeable effect on accident rates.

\subsection{Crash Severity}
The correlation between pavement conditions and other characteristics and crash severity has also been extensively examined in the literature. According to \citet{li2013ImpactPavementConditions}, smoother pavements, as indicated by a higher IRI or ride score, are associated with increased crash severity, including fatal accidents. Additionally, \citet{buddhavarapu2013InfluencePavementCondition} emphasized that fatal crashes are more likely to occur on smoother pavements.

The overall condition of the pavement plays a complex role in crash dynamics. Poor pavement conditions are generally linked with more severe crashes, particularly impacting passenger vehicles more significantly than commercial vehicles. However, the correlation is not consistently observed, as very poor distress and condition scores do not always result in heightened crash severity. However, excellent pavement conditions can sometimes lead to severe outcomes \citep{li2013ImpactPavementConditions}. \citet{lee2015effects} added that the impact of poor pavement condition varies by speed level and vehicle involvement, with severity increasing for multi-vehicle crashes across all speed levels and for single-vehicle crashes on high-speed roads, while it decreases on low-speed roads as pavement condition worsens.

Various additional roadway features also influence crash severity. Wider shoulders on two-lane curves are associated with higher fatality rates in crashes. Geographical location is another critical factor, with rural areas experiencing more severe and fatal crashes compared to urban settings. The presence of vertical or horizontal curves, especially those with a downward gradient, increases the likelihood of fatal outcomes \citep{buddhavarapu2013InfluencePavementCondition}. Weather conditions and road surface types contribute to crash severity variations. \citet{lee2015effects} and \citet{becker2022weather} observed that wet surfaces tend to lower severity across all crash types, challenging some common perceptions about wet weather driving. Similarly, the skid resistance, particularly on two-lane horizontal curves, shows a weak correlation with crash severity \citep{buddhavarapu2013InfluencePavementCondition}. Intersections have an impact on crash severity depending on the speed and number of vehicles involved. At intersections, crash severity tends to decrease for single-vehicle incidents on low- and medium-speed roads but increases for multi-vehicle accidents \citep{sharafeldin2022investigating}. Regarding the influence of speed limits, they generally mitigate crash severity except in cases of single-vehicle accidents on low- and medium-speed roads \citep{lee2015effects}. Temporal factors such as the time of day also play a significant role. Daytime crashes are more likely to be fatal compared to nighttime incidents, and crashes tend to be more severe in darker conditions for multi-vehicle collisions on medium-speed roads, with a decrease in severity noted for similar crashes on high-speed roads \citep{zhang2022random}. Lastly, the percentage of trucks involved in a crash also affects severity, with higher truck traffic increasing the risk of fatalities but reducing severity in multi-vehicle crashes on lower speed roads \citep{buddhavarapu2013InfluencePavementCondition, lee2015effects}.

\subsection{Research Motivation and Objective}

As discussed earlier, the impact of pavement conditions and other road-related variables on crash frequency and severity is complex and varies significantly across different studies. Some research has identified positive correlations, while others have found negative associations, with these findings often differing by geographic location. The complexity of this relationship, compounded by the interplay of geometric, environmental, and operational factors, highlights the need for more granular and localized analysis. To further investigate these discrepancies, this study utilizes data from the Iowa Department of Transportation (DOT) to analyze the effects of these variables and uncover new insights. The objective of this study is to quantify the influence of pavement performance indicators, including roughness, friction, rutting, and surface distress, on crash frequency and severity. 

\section{Methodology}

This section outlines the mathematical models employed to analyze the relationship between pavement performance and road safety, focusing on crash frequency and severity. The primary models include the negative binomial regression for crash frequency and the ordered probit regression for crash severity, supplemented by the Tukey Honest Significant Difference (HSD) test and random forest models.

\subsection{Negative Binomial Regression for Crash Frequency}

The negative binomial regression model is used to model crash frequency, which is suitable for count data exhibiting overdispersion \citep{wu2014analysis}. For each road segment \( i \), the number of crashes \( Y_i \) is assumed to follow a negative binomial distribution, denoted as \( Y_i \sim \text{NegBin}(\mu_i, \theta) \), with mean \( \mu_i \) and dispersion parameter \( \theta \). The expected crash frequency \( \mu_i \) is formulated as:
\begin{equation}
\log(\mu_i) = \beta_0 + \beta_1 x_{i1} + \beta_2 x_{i2} + \dots + \beta_p x_{ip}
\end{equation}
where \( \mu_i \) represents the expected number of crashes for segment \( i \), \( \beta_0 \) is the intercept, \( \beta_1, \beta_2, \dots, \beta_p \) are the coefficients, and \( x_{i1}, x_{i2}, \dots, x_{ip} \) are the predictor variables for segment \( i \). The connection between \( Y_i \) and \( \mu_i \) is defined by the negative binomial distribution, where \( Y_i \) has mean \( \mu_i \) and variance \( \text{Var}(Y_i) = \mu_i + \frac{\mu_i^2}{\theta} \), with \( \theta \) controlling overdispersion \citep{cameron2013regression}.

\subsection{Ordered Probit Regression for Crash Severity}

The ordered probit regression model is applied to model crash severity, an ordinal outcome. It assumes a latent continuous variable \( Y^* \) underlying the observed severity levels, expressed as:
\begin{equation}
Y^* = X^T \beta + \epsilon
\end{equation}
where \( X \) is the vector of explanatory variables, \( \beta \) is the vector of coefficients, and \( \epsilon \) is an error term following a standard normal distribution. The observed severity level \( Y \) is determined by thresholds \( \tau_k \), and the probability of observing a specific severity level \( k \) is:
\begin{equation}
P(Y = k) = \Phi(\tau_k - X^T \beta) - \Phi(\tau_{k-1} - X^T \beta)
\end{equation}
where \( \Phi \) is the cumulative distribution function of the standard normal distribution, and \( \tau_{k-1} < \tau_k \) are the estimated thresholds \citep{borooah2002logit}. 

\subsection{Mean Comparison Using Tukey Test}

The Tukey HSD test is utilized for comparing means across categories. The test statistic for comparing two means \( \bar{Y}_i \) and \( \bar{Y}_j \) is given by:
\begin{equation}
q = \frac{\bar{Y}_i - \bar{Y}_j}{\text{SE}}
\end{equation}
where SE is the standard error of the difference between the means. This statistic is compared to a critical value from the studentized range distribution to determine statistical significance \citep{hsu1996multiple}.

\subsection{Random Forest Models}

Random forest models are employed to assess feature importance for both crash frequency and severity. The importance of a feature \( x_j \) is calculated using the mean decrease in impurity (MDI):
\begin{equation}
\text{Importance}(x_j) = \frac{1}{N_T} \sum_{t=1}^{N_T} \Delta I(t, x_j)
\end{equation}
where \( N_T \) is the number of trees in the forest, and \( \Delta I(t, x_j) \) is the decrease in impurity when splitting on feature \( x_j \) in tree \( t \). This measure quantifies the contribution of each feature to the model's predictive performance \citep{breiman2001random}.

In this study, we employ both a machine learning approach (random forest model) and econometric models (negative binomial regression and ordered probit regression) to analyze the relationship between pavement performance and road safety. These two groups of models serve fundamentally different purposes: The random forest model is designed for prediction. Its non-parametric, ensemble-based structure captures complex, non-linear relationships without assuming a specific data distribution. It also ranks feature importance, making it ideal for forecasting crash outcomes. In contrast, econometric models are tailored for inference. They rely on parametric assumptions and estimate interpretable coefficients and test statistical significance. This helps reveal causal or associative effects of pavement conditions on crash metrics. Because these models have different goals, prediction versus inference, their performance can't be directly compared. Each model provides unique insights, and using both allows for a more comprehensive understanding of the link between pavement conditions and road safety.

\section{Dataset}\label{dataset}

In this study, we examined traffic crashes that occurred in Iowa during the year 2022. Our data collection and processing efforts comprised the following three main steps: 1) Crash data preparation, 2) Pavement inventory data preparation, and 3) Integrating crash and pavement data.

\subsection{Crash Dataset}\label{crash-dataset}

The crash dataset was sourced from the Iowa Department of Transportation (DOT) \citep{IowaDOT_RoadwayData}. 
% Table~\ref{tab_crashrate} outlines the variables included in the dataset, along with their meanings and potential values. 
The dataset cover a broad spectrum of information, including the crash timestamp (TIMESTR), and crash location coordinates (XCOORD, YCOORD). Details regarding specific conditions at the time of the crash are also recorded, such as the location of the initial harmful event (LOCFSTHRM), involvement of drugs or alcohol (DRUGALC), and environmental contributing factors (ECNTCRC). Light conditions (LIGHT), surface conditions (CSRFCND), weather (WEATHER), roadway-related contributing factors (RCNTCRC), road pavement status (PAVED), whether the crash occurred within a work zone (WZRELATED), and the severity of the crash (CSEV) are included.

\begin{figure}[H]
    \centering
    \includegraphics[width=\textwidth]{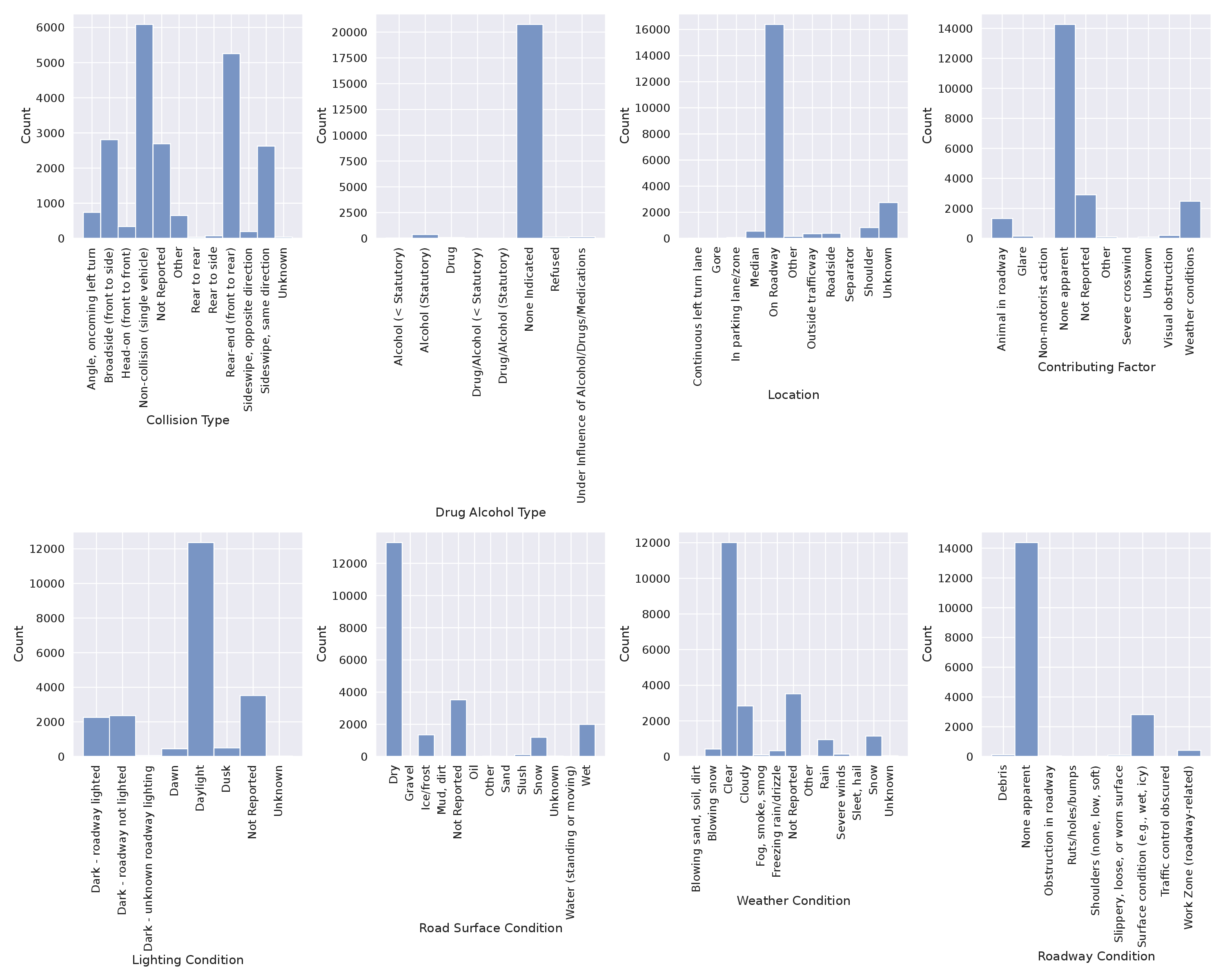}
    \caption{Histogram of the Crash Dataset}\label{fig_CR_histogram}
\end{figure}

Figure~\ref{fig_CR_histogram} displays histograms showing the distribution of key variables in the crash dataset. The Collision\_Type histogram indicates that the most frequent crash types are non-collision (single vehicle) and rear-end (front-to-rear) crashes. For Drug\_Alcohol\_Type, the majority of cases report no involvement of drugs or alcohol. The Location histogram shows that most crashes occur on roadways, with a notable portion recorded at unknown locations. Lighting\_Condition reveals that the majority of crashes happen during daylight hours. In Road\_Surface\_Condition, dry roads are the most common, followed by wet and snow-covered surfaces. The Weather\_Condition histogram highlights clear weather as the most prevalent, though ``Not Reported'' also appears frequently. Lastly, Roadway\_Condition shows that ``None apparent'' is the most reported condition, followed by ``Surface condition (e.g., wet, icy).''

\subsection{Pavement Dataset}\label{pavement-dataset}

\begin{figure}[H]
    \centering
    \includegraphics[width=\textwidth]{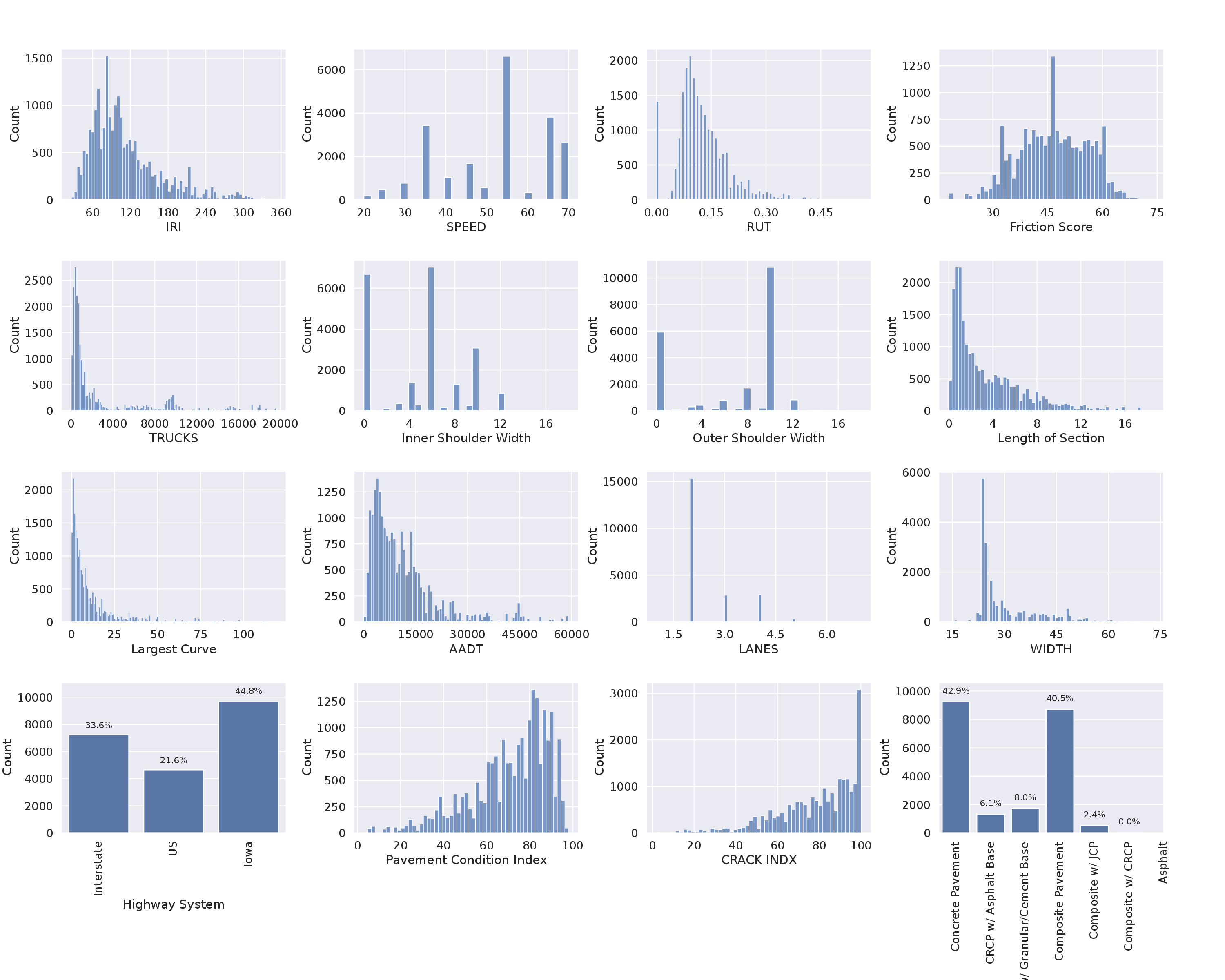}
    \caption{Histogram of Pavement Dataset}\label{fig_pavement_histogram}
\end{figure}

The pavement inventory dataset sourced from Iowa DOT provides various aspects of pavement conditions. 
% Table~\ref{tab_pavementdataset} summarizes these statistics, while 
Figure~\ref{fig_pavement_histogram} illustrates the histograms of the variables in the pavement dataset. The IRI measures pavement smoothness, with a mean of 112 and a standard deviation of 55.2. The histogram shows that most roads have an IRI below 200, but a few roads are significantly rougher. Speed limits (SPEED) average 52 miles per hour (mph), ranging from 20 to 70 mph, which reflects a mix of urban and highway roads. Most roads have shallow ruts, as indicated by the histogram, which shows a concentration around 0 to 0.2 inches, with very few roads having ruts deeper than 0.4 inches. The friction score (FRICT) are mostly concentrated between 40 and 60. Average annual daily traffic (AADT) varies greatly, with a mean of 11,133 vehicles and a high standard deviation. The inner shoulder width (ISHLDWID) averages 4.92 feet, with values ranging from 0 to 18 feet. Similarly, the outer shoulder width (OSHLDWID) has a higher average of 6.70 feet and varies within the same 0 to 18 feet range. The largest degree of curve (LARGEST CURVE) averages 10.83 degrees, but the range extends up to 118.85 degrees, which reflects significant differences in road alignment and curvature complexity across segments. The length of road segments (PMIS LENGTH) shows considerable variability, with an average of 3.42 miles and a spread from as short as 0.01 miles to as long as 18.59 miles. The number of lanes (LANES) histogram indicates that most roads have 2 lanes, with fewer roads having 3, 4, or more lanes. Pavement width histogram shows a concentration around 20 to 30 feet, with some roads wider than 60 feet. The dataset includes roads from three different systems (SYSTEM): Interstate, US highways, and Iowa state highways. The Pavement Condition Index (PCI\_2) averages 70.78, showing generally good conditions but with variability. The cracking index (CRACK\_INDX) histogram shows a wide range, with most roads scoring above 60, indicating moderate to low cracking. The dataset includes multiple pavement types (PAVTYP), ranging from concrete and asphalt to various composite types.

\subsection{Crash and Pavement Data Integration}\label{crash-and-pavement-data-integration}

\begin{figure}[H]
    \centering
    \includegraphics[width=0.8\textwidth, trim={50 150 50 150}, clip]{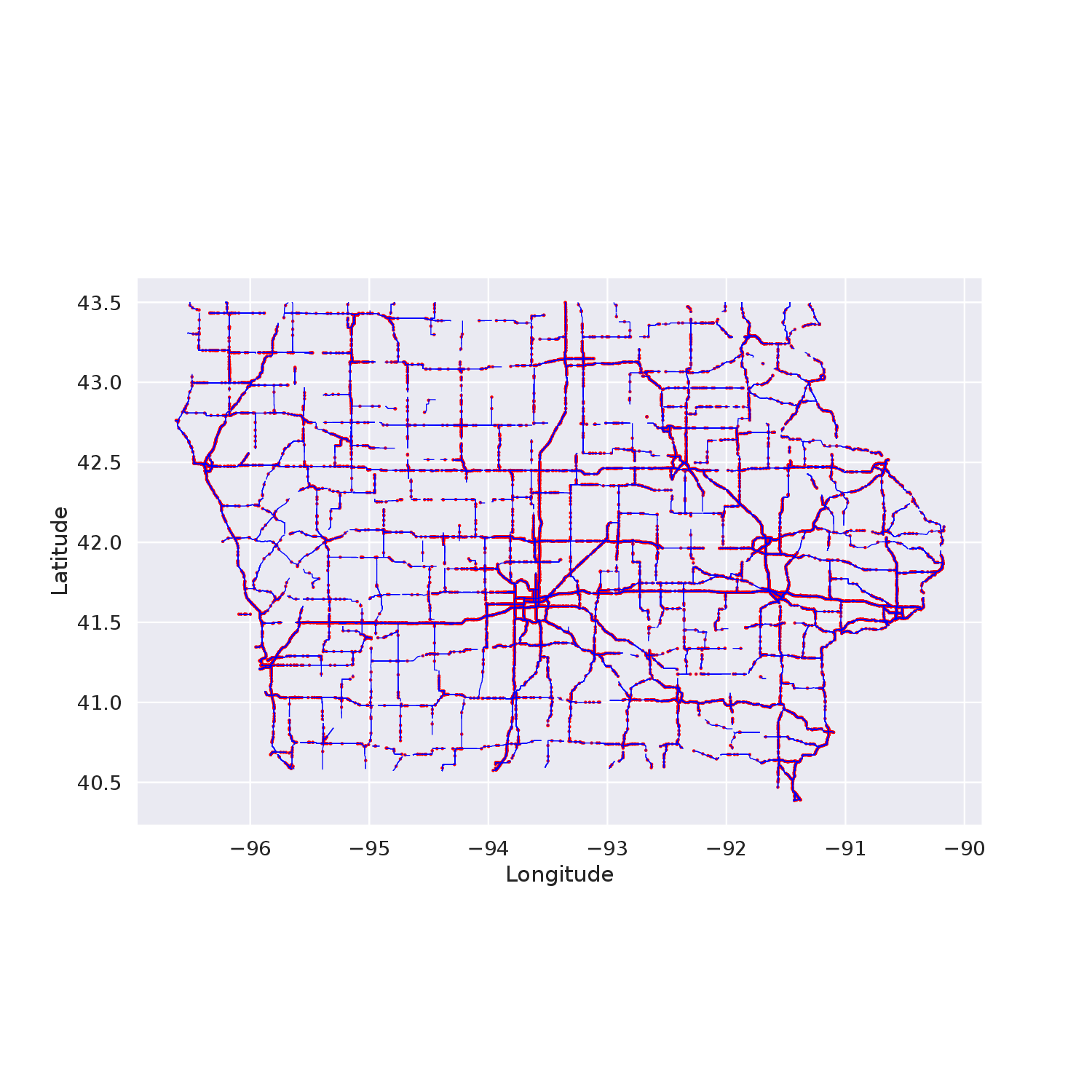}
    \caption{Crash Map}\label{fig:fig_crashmap}
\end{figure}

The crash data was transformed into a Geodata frame, associating each crash record with its geographical coordinates. Meanwhile, the pavement dataset was parsed to extract LineString geometries representing road segments. To facilitate spatial analysis, the LineString geometries were buffered to create zones around road segments, representing areas where pavement conditions might influence crash incidents. A spatial join operation was then performed between the buffered road segments and the Geodata Frame of crash data to identify points (crash locations) near the road segments. Finally, there are 21,536 crash points which occurred near to the Road segments with buffer zone (Figure \ref{fig:fig_crashmap}).

\section{Mean Comparison of Crash Rate and Crash Severity}\label{mean-comparison-of-crash-rate-and-crash-severity}

To compare the impact of different groups of pavement conditions on the crash rate and crash severity, bar charts were generated to demonstrate the general relationship between these factors and various pavement condition metrics. The Tukey test was used to determine whether the differences in crash rates between various pavement condition categories were statistically significant. By comparing each pair of condition groups, the specific ranges of road condition that had a more pronounced effect on crash rates or severity can be identified. It's worth noting that not all tables contain the same number of scenarios. This discrepancy arises when a scenario lacks sufficient data points for meaningful statistical comparison, leading to its removal from the table.

\subsection{Crash Rate Across Pavement Conditions}\label{crash-rate}

In this case study, the crash rate is defined as:

\begin{equation}\label{equ_cr}
    R=\frac{C \times 100,000,000}{V*L*365}
\end{equation}

\noindent where,\\
R = crash rate for the road segment (crashes per 100 million vehicle-miles of travel); \\
C = total number of crashes in 2022; \\
V = number of vehicles per day; \\
L = length of the pavement segment.

Based on previous studies and the distribution of pavement index values, the pavement condition index was defined into different groups. For instance, the International Roughness Index (IRI) scores were categorized into three groups: Good for scores from 0 to 95, Fair for scores between 95 and 170, and Poor for scores exceeding 170 \citep{fhwa_pavement_2024}. We categorized rutting depth into three categories: Good (0 to 1/2 inch), Fair (1/2 inch to 3/4 inch), and Poor (over 3/4 inch). The Pavement Condition Index (PCI) is classified into three categories: Good [70-100], Fair [40-70), and Poor [0-40) \citep{han2023pcier}. Friction is defined into three groups as Good [51-75], Fair [26-51), and Poor [1-26). The Crack Index is divided into three groups: Good (80-100], Fair (60-80], and Poor [0-60].

% \hl{definition for FRICT and Crack Index}.

\begin{figure}[H]
    \centering
    \includegraphics[width=\textwidth]{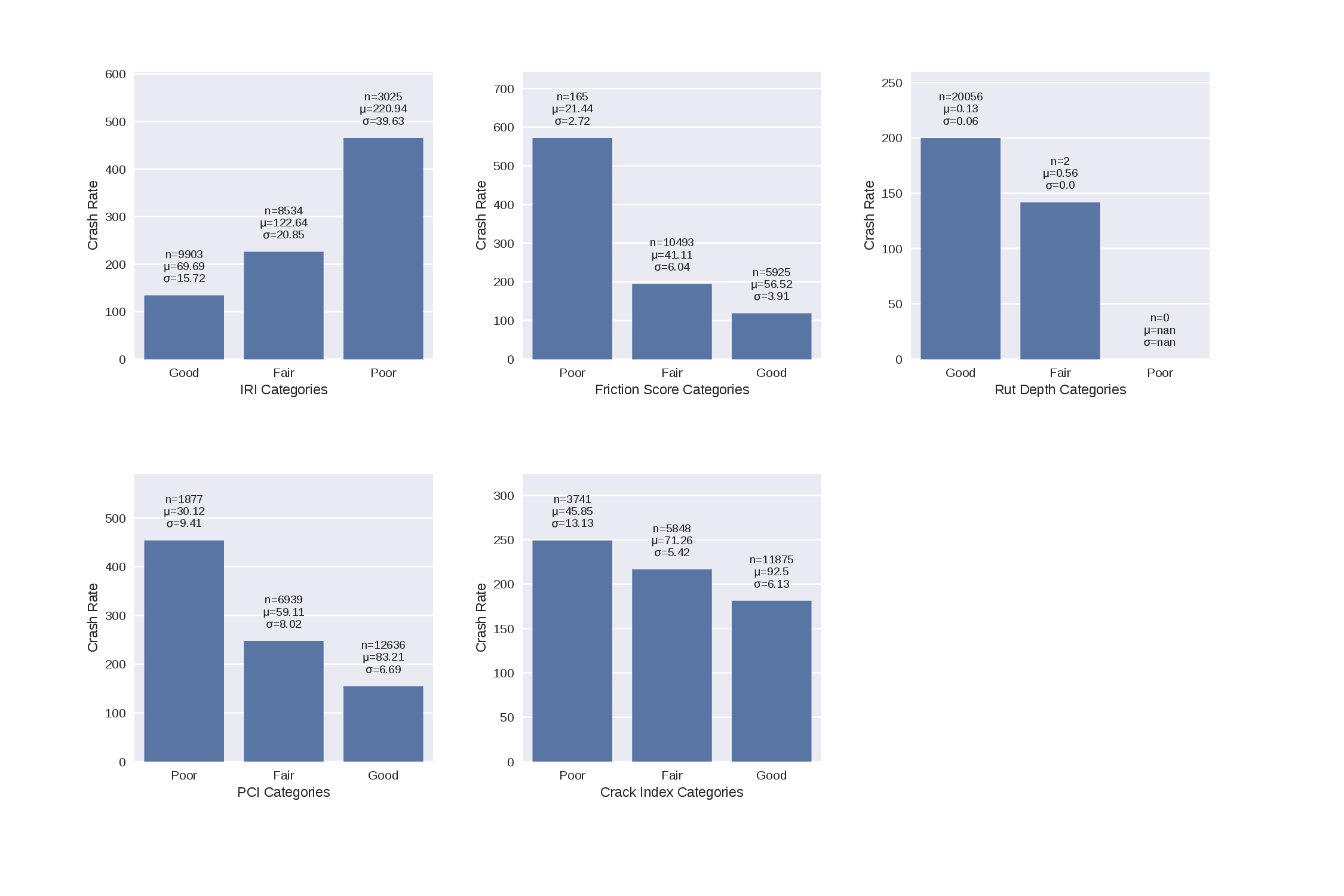}
    \caption{Mean Crash Rates vs. Pavement Conditions}
    \caption*{\small Note: $n$ denotes the number of data points in each category, $\mu$ is the mean pavement condition index, and $\sigma$ is the standard deviation.}
    \label{fig_cr_pavement}
\end{figure}

Figure~\ref{fig_cr_pavement} illustrates the relationship between various pavement condition metrics and crash rates. The value of each pavement condition metric is equally divided into three groups, and the bar charts illustrate the mean crash rates for each group. The IRI plot shows that crash rates increase with higher IRI values, indicating rougher roads are associated with more crashes. For the friction score (FRICT), there is an inverse relationship with crash rates. Higher friction scores are associated with lower crash rates. The Rut Depth (RUT) plot indicates that shallower rutting depths are associated with higher crash rates, which contradicts the common belief that better pavement conditions correlate with fewer crashes. This counterintuitive finding may be explained by the tendency of drivers to slow down in the presence of severe rutting, thereby reducing the crash rate. Additionally, the distribution of data may play a role, as there were very few data points in the Poor category. The Pavement Condition Index (PCI\_2) shows a clear inverse relationship with crash rates. For Cracking Index (CRACK\_INDX), better pavement is associated with fewer crash rate but the difference is not that significant. Based on these observations and the correlation between PCI\_2, CRACK\_INDEX, and IRI, we have chosen IRI and FRICT for further analysis using the Tukey Test.

% \hl{1x5 figures}

\subsubsection{Crash Rate Analysis by IRI}\label{iri}

% Table \ref{tab_cr_IRI} shows the mean comparison results using the Tukey test. The overall crash rate significantly increases from ``good'' to ``poor'' pavement conditions. The Tukey test p-values confirm that the differences between ``good,'' ``fair,'' and ``poor'' categories are highly significant.

Table \ref{tab_cr_IRI} presents the mean comparison results using the Tukey test, showing that the overall crash rate significantly increases from ``good" to ``poor" pavement conditions. The Tukey test p-values confirm that the differences between the ``good", ``fair", and ``poor" categories are highly significant. This difference exists in most scenarios shown in Table \ref{tab_cr_IRI}. However, the difference is not significant in the following collision type scenarios: sideswipe (opposite), rear-to-rear collision, and rear-to-side collision, possibly because these types of collisions are often more influenced by driver behavior, such as lane changes or sudden stops, rather than pavement condition. For weather conditions, the difference in crash rates across different IRI categories is not significant in scenarios involving fog/smoke/smog and blowing snow, perhaps because the primary factor influencing crashes in poor visibility conditions is the limited visibility rather than the road surface condition. For road surface conditions, the differences between categories are not significant in scenarios involving slush, sand, and gravel, likely because these surface conditions introduce additional hazards that overshadow the influence of pavement roughness. For the location of the first harmful event, the difference is not significant in scenarios occurring in the median, parking lane/zone, and continuous left turn lane, as crashes in these locations may be more related to specific traffic maneuvers, such as making turns or parking, where the pavement condition has less influence compared to other factors like traffic density and driver attentiveness. For roadway-related contributing circumstances, the difference is not significant in scenarios involving debris, ruts/holes/bumps, and obstructions in the roadway, as the presence of specific obstacles or hazards likely plays a more critical role in crash occurrence than the overall roughness of the pavement. The difference between IRI categories is also not significant in many scenarios related to alcohol and in scenarios involving visual obstructions, glare, and animals in the roadway, as these crashes are likely heavily influenced by external factors such as impaired driving or sudden visual obstructions, which can be severe regardless of pavement condition.

\begin{table}[H]
    \caption{Mean Crash Rates by IRI Groups}\label{tab_cr_IRI}
    \footnotesize
    \setlength{\tabcolsep}{3pt}
    \renewcommand{\arraystretch}{0.85}
    \centering
    \begin{tabular}{|l|l|ccc|ccc|}
        \hline
        & & \multicolumn{3}{c|}{Avg. Crash Rate} & \multicolumn{3}{c|}{Tukey p-value} \\
        & Scenario & Good & Fair & Poor & G-F & F-P & G-P \\
        \hline
        & \rule{0pt}{10pt}Overall Crash Rate & 134 & 226 & 466 & 0.0000 & 0.0000 & 0.0000 \\
        \hline
        \multirow{9}{*}{\rotatebox{90}{Collision Type}} 
        & \rule{0pt}{10pt}Single Vehicle & 56 & 72 & 119 & 0.0004 & 0.0000 & 0.0000 \\
        & \rule{0pt}{10pt}Head On & 33 & 50 & 61 & 0.1024 & 0.4915 & 0.0242 \\
        & \rule{0pt}{10pt}Rear End & 56 & 117 & 201 & 0.0000 & 0.0000 & 0.0000 \\
        & \rule{0pt}{10pt}Angle Left Turn & 53 & 81 & 91 & 0.0134 & 0.6116 & 0.0030 \\
        & \rule{0pt}{10pt}Broadside & 83 & 130 & 205 & 0.0001 & 0.0000 & 0.0000 \\
        & \rule{0pt}{10pt}Sideswipe (Same) & 36 & 69 & 142 & 0.0001 & 0.0000 & 0.0000 \\
        & \rule{0pt}{10pt}Sideswipe (Opp.) & 40 & 44 & 35 & 0.9280 & 0.7378 & 0.8886 \\
        & \rule{0pt}{10pt}Rear to Rear & 49 & 83 & 73 & 0.6333 & 0.9606 & 0.8155 \\
        & \rule{0pt}{10pt}Rear to Side & 46 & 61 & 83 & 0.7533 & 0.7346 & 0.4422 \\
        \hline
        \multirow{7}{*}{\rotatebox{90}{Weather Condition}} 
        & \rule{0pt}{10pt}Clear & 88 & 171 & 362 & 0.0000 & 0.0000 & 0.0000 \\
        & \rule{0pt}{10pt}Cloudy & 41 & 78 & 131 & 0.0000 & 0.0000 & 0.0000 \\
        & \rule{0pt}{10pt}Freezing Rain & 20 & 36 & 52 & 0.0081 & 0.0090 & 0.0000 \\
        & \rule{0pt}{10pt}Rainy & 32 & 50 & 117 & 0.1948 & 0.0000 & 0.0000 \\
        & \rule{0pt}{10pt}Snow & 23 & 39 & 73 & 0.0000 & 0.0000 & 0.0000 \\
        & \rule{0pt}{10pt}Fog/Smoke & 26 & 44 & 52 & 0.3984 & 0.9293 & 0.4844 \\
        & \rule{0pt}{10pt}Blowing Snow & 19 & 27 & 35 & 0.0481 & 0.5742 & 0.0896 \\
        \hline
        \multirow{7}{*}{\rotatebox{90}{Road Surface}} 
        & \rule{0pt}{10pt}Dry & 95 & 183 & 384 & 0.0000 & 0.0000 & 0.0000 \\
        & \rule{0pt}{10pt}Wet & 38 & 64 & 131 & 0.0004 & 0.0000 & 0.0000 \\
        & \rule{0pt}{10pt}Ice/Frost & 25 & 40 & 55 & 0.0000 & 0.0388 & 0.0000 \\
        & \rule{0pt}{10pt}Snow & 25 & 43 & 84 & 0.0000 & 0.0000 & 0.0000 \\
        & \rule{0pt}{10pt}Slush & 27 & 59 & 40 & 0.3220 & 0.8055 & 0.9047 \\
        & \rule{0pt}{10pt}Sand & 46 & 10 & 72 & 0.9362 & 0.8393 & 0.9532 \\
        & \rule{0pt}{10pt}Gravel & 31 & 31 & 21 & 0.9988 & 0.9433 & 0.9321 \\
        \hline
        \multirow{7}{*}{\rotatebox{90}{Location}} 
        & \rule{0pt}{10pt}On Roadway & 107 & 206 & 445 & 0.0000 & 0.0000 & 0.0000 \\
        & \rule{0pt}{10pt}Shoulder & 24 & 39 & 58 & 0.0069 & 0.2688 & 0.0144 \\
        & \rule{0pt}{10pt}Off Traffic & 25 & 47 & 121 & 0.4663 & 0.0112 & 0.0005 \\
        & \rule{0pt}{10pt}Median & 14 & 17 & 27 & 0.2174 & 0.3856 & 0.1699 \\
        & \rule{0pt}{10pt}Roadside & 22 & 34 & 49 & 0.0894 & 0.3748 & 0.0343 \\
        & \rule{0pt}{10pt}Parking Lane & 48 & 55 & 112 & 0.9657 & 0.2229 & 0.1642 \\
        & \rule{0pt}{10pt}Left Turn Lane & 29 & 43 & 35 & 0.8583 & 0.8410 & 0.9732 \\
        \hline
        \multirow{6}{*}{\rotatebox{90}{Road Factors}} 
        & \rule{0pt}{10pt}Wet/Icy & 35 & 56 & 101 & 0.0000 & 0.0000 & 0.0000 \\
        & \rule{0pt}{10pt}Debris & 10 & 22 & 33 & 0.0941 & 0.6084 & 0.1128 \\
        & \rule{0pt}{10pt}Work Zone & 39 & 40 & 84 & 0.9992 & 0.0375 & 0.0282 \\
        & \rule{0pt}{10pt}Slippery/Worn & 25 & 40 & 265 & 0.9694 & 0.0438 & 0.0241 \\
        & \rule{0pt}{10pt}Ruts/Holes & 12 & 9 & 18 & 0.9553 & 0.8238 & 0.8993 \\
        & \rule{0pt}{10pt}Obstruction & 14 & 37 & 23 & 0.1753 & 0.8646 & 0.9326 \\
        \hline
        \multirow{6}{*}{\rotatebox{90}{Drug/Alcohol}} 
        & \rule{0pt}{10pt}Only Drugs & 13 & 23 & 39 & 0.0613 & 0.0982 & 0.0017 \\
        & \rule{0pt}{10pt}Alcohol (Stat.) & 24 & 45 & 77 & 0.0364 & 0.0609 & 0.0003 \\
        & \rule{0pt}{10pt}Under Influence & 28 & 38 & 79 & 0.6040 & 0.0522 & 0.0074 \\
        & \rule{0pt}{10pt}Drug/Alc ($<$Stat.) & 32 & 51 & 78 & 0.6394 & 0.6070 & 0.2650 \\
        & \rule{0pt}{10pt}Drug/Alc (Stat.) & 25 & 25 & 20 & 0.9998 & 0.9841 & 0.9736 \\
        & \rule{0pt}{10pt}Refused & 23 & 38 & 56 & 0.2429 & 0.3890 & 0.0375 \\
        \hline
        \multirow{5}{*}{\rotatebox{90}{Lighting}} 
        & \rule{0pt}{10pt}Daylight & 87 & 179 & 381 & 0.0000 & 0.0000 & 0.0000 \\
        & \rule{0pt}{10pt}Dusk & 23 & 38 & 51 & 0.0011 & 0.1155 & 0.0000 \\
        & \rule{0pt}{10pt}Dark (Lighted) & 52 & 93 & 179 & 0.0001 & 0.0000 & 0.0000 \\
        & \rule{0pt}{10pt}Dark (Unlit) & 35 & 40 & 55 & 0.2092 & 0.0615 & 0.0072 \\
        & \rule{0pt}{10pt}Dawn & 24 & 45 & 77 & 0.0364 & 0.0609 & 0.0003 \\
        \hline
        \multirow{5}{*}{\rotatebox{90}{Environment}} 
        & \rule{0pt}{10pt}Weather & 31 & 47 & 80 & 0.0000 & 0.0000 & 0.0000 \\
        & \rule{0pt}{10pt}Visual Obst. & 47 & 57 & 64 & 0.7445 & 0.9073 & 0.5478 \\
        & \rule{0pt}{10pt}Glare & 38 & 47 & 45 & 0.6971 & 0.9942 & 0.8919 \\
        & \rule{0pt}{10pt}Animal & 36 & 39 & 54 & 0.7401 & 0.3472 & 0.2054 \\
        & \rule{0pt}{10pt}None Apparent & 97 & 196 & 419 & 0.0000 & 0.0000 & 0.0000 \\
        \hline
    \end{tabular}
\end{table}

\subsubsection{Crash Rate Analysis by Friction}\label{friction}

Table~\ref{tab_cr_friction} presents the results of mean comparisons using the Tukey test to examine the relationship between friction scores and crash rates. The overall crash rate significantly increases from ``good" to ``poor" pavement conditions. The differences between different friction categories are significant in about half of the scenarios. Notably, they are not significant in scenarios such as single vehicle collision, head-on (front to front), angle oncoming left turn, rear-to-rear collision, freezing rain/drizzle, rainy, blowing snow, ice/frost surface condition, outside traffic way, in parking lane/zone, slippery loose or worn, alcohol (statutory), dusk lighting condition, dark road not lighted, dawn, and glare scenarios. This may be because in these scenarios, factors other than pavement friction, such as driver behavior, environmental conditions, and specific traffic dynamics, play a more dominant role in crash occurrences. For example, in poor visibility conditions (dusk, dawn, dark road not lighted, and glare), visibility issues may overshadow the impact of pavement friction. In weather-related scenarios (freezing rain/drizzle, rainy, blowing snow, ice/frost), the primary factor influencing crashes could be the challenging driving conditions rather than the friction of the pavement. In collision types like single vehicle collision, head-on, angle oncoming left turn, and rear-to-rear, driver errors and specific maneuver dynamics might be more critical than pavement friction. Similarly, in areas outside traffic ways or parking zones, crashes might be more influenced by specific situational factors and driver actions than by pavement conditions. In scenarios involving alcohol, impaired driving likely plays a more significant role in crashes than the friction of the pavement.

\begin{table}[H]
    \caption{Mean Crash Rates by Friction Score Groups}\label{tab_cr_friction}
    \footnotesize
\setlength{\tabcolsep}{3pt}
\renewcommand{\arraystretch}{0.85}    
    \centering
    \begin{tabular}{|l|l|ccc|ccc|}
        \hline
        Category & Scenario & Good & Fair & Poor & G-F & F-P & G-P \\
        \hline
        \rule{0pt}{10pt}Overall & Overall Crash Rate & 118 & 194 & 571 & 0.0000 & 0.0000 & 0.0000 \\
        \hline
        \multirow{7}{*}{Collision Type} 
        & \rule{0pt}{10pt}Single Vehicle & 60 & 57 & 103 & 0.3128 & 0.5115 & 0.2596 \\
        & \rule{0pt}{10pt}Head On & 39 & 44 & 50 & 0.9281 & 0.8571 & 0.9788 \\
        & \rule{0pt}{10pt}Rear End & 47 & 95 & 180 & 0.0000 & 0.1488 & 0.0108 \\
        & \rule{0pt}{10pt}Angle Left Turn & 46 & 66 & 84 & 0.0807 & 0.8091 & 0.4218 \\
        & \rule{0pt}{10pt}Broadside & 59 & 101 & 235 & 0.0000 & 0.0155 & 0.0009 \\
        & \rule{0pt}{10pt}Sideswipe (Same) & 30 & 47 & 76 & 0.0001 & 0.4071 & 0.1105 \\
        & \rule{0pt}{10pt}Rear to Rear & 20 & 66 & 46 & 0.8744 & 0.9745 & 0.9779 \\
        \hline
        \multirow{6}{*}{Weather Condition} 
        & \rule{0pt}{10pt}Clear & 71 & 137 & 390 & 0.0000 & 0.0000 & 0.0000 \\
        & \rule{0pt}{10pt}Cloudy & 42 & 59 & 172 & 0.0006 & 0.0008 & 0.0001 \\
        & \rule{0pt}{10pt}Freezing Rain & 26 & 26 & 75 & 0.9993 & 0.0977 & 0.1023 \\
        & \rule{0pt}{10pt}Rainy & 35 & 37 & 34 & 0.9176 & 0.9966 & 1.0000 \\
        & \rule{0pt}{10pt}Snow & 23 & 34 & 54 & 0.0030 & 0.5284 & 0.2351 \\
        & \rule{0pt}{10pt}Blowing Snow & 23 & 21 & 57 & 0.7921 & 0.4414 & 0.4897 \\
        \hline
        \multirow{4}{*}{Road Surface} 
        & \rule{0pt}{10pt}Dry & 75 & 148 & 447 & 0.0000 & 0.0000 & 0.0000 \\
        & \rule{0pt}{10pt}Wet & 35 & 50 & 89 & 0.0014 & 0.2403 & 0.0650 \\
        & \rule{0pt}{10pt}Ice/Frost & 30 & 32 & 48 & 0.8588 & 0.7879 & 0.7460 \\
        & \rule{0pt}{10pt}Snow & 27 & 37 & 51 & 0.0139 & 0.7662 & 0.4439 \\
        \hline
        \multirow{4}{*}{Location} 
        & \rule{0pt}{10pt}On Roadway & 86 & 173 & 604 & 0.0000 & 0.0000 & 0.0000 \\
        & \rule{0pt}{10pt}Shoulder & 34 & 23 & 138 & 0.0399 & 0.0605 & 0.1014 \\
        & \rule{0pt}{10pt}Off Traffic & 29 & 28 & 60 & 0.9659 & 0.3943 & 0.4235 \\
        & \rule{0pt}{10pt}Parking Lane & 55 & 58 & 23 & 0.9962 & 0.9483 & 0.9608 \\
        \hline
        \multirow{2}{*}{Road Factors} 
        & \rule{0pt}{10pt}Wet/Icy & 34 & 51 & 57 & 0.0001 & 0.9676 & 0.6452 \\
        & \rule{0pt}{10pt}Slippery/Worn & 27 & 28 & 13 & 0.9826 & 0.9023 & 0.9210 \\
        \hline
        \multirow{3}{*}{Drug/Alcohol} 
        & \rule{0pt}{10pt}Only Drugs & 21 & 18 & 23 & 0.7546 & 0.9631 & 0.9962 \\
        & \rule{0pt}{10pt}Alcohol (Stat.) & 30 & 29 & 18 & 0.9766 & 0.9081 & 0.8924 \\
        & \rule{0pt}{10pt}Refused & 19 & 26 & 14 & 0.6386 & 0.9407 & 0.9913 \\
        \hline
        \multirow{5}{*}{Lighting} 
        & \rule{0pt}{10pt}Daylight & 71 & 145 & 472 & 0.0000 & 0.0000 & 0.0000 \\
        & \rule{0pt}{10pt}Dusk & 27 & 29 & 44 & 0.8767 & 0.7617 & 0.7044 \\
        & \rule{0pt}{10pt}Dark (Lighted) & 42 & 67 & 97 & 0.0015 & 0.6468 & 0.2514 \\
        & \rule{0pt}{10pt}Dark (Unlit) & 37 & 37 & 69 & 0.9993 & 0.7895 & 0.7907 \\
        & \rule{0pt}{10pt}Dawn & 28 & 29 & 17 & 0.9943 & 0.9088 & 0.9178 \\
        \hline
        \multirow{3}{*}{Environment} 
        & \rule{0pt}{10pt}Weather & 30 & 44 & 34 & 0.0000 & 0.8640 & 0.9784 \\
        & \rule{0pt}{10pt}Glare & 45 & 39 & 18 & 0.8886 & 0.8687 & 0.8050 \\
        & \rule{0pt}{10pt}None Apparent & 74 & 160 & 517 & 0.0000 & 0.0000 & 0.0000 \\
        \hline
    \end{tabular}
\end{table}

\subsection{Crash Severity Across Pavement Conditions}\label{crash-severity-1}

\begin{figure}[H]
    \centering
    \includegraphics[width=\textwidth]{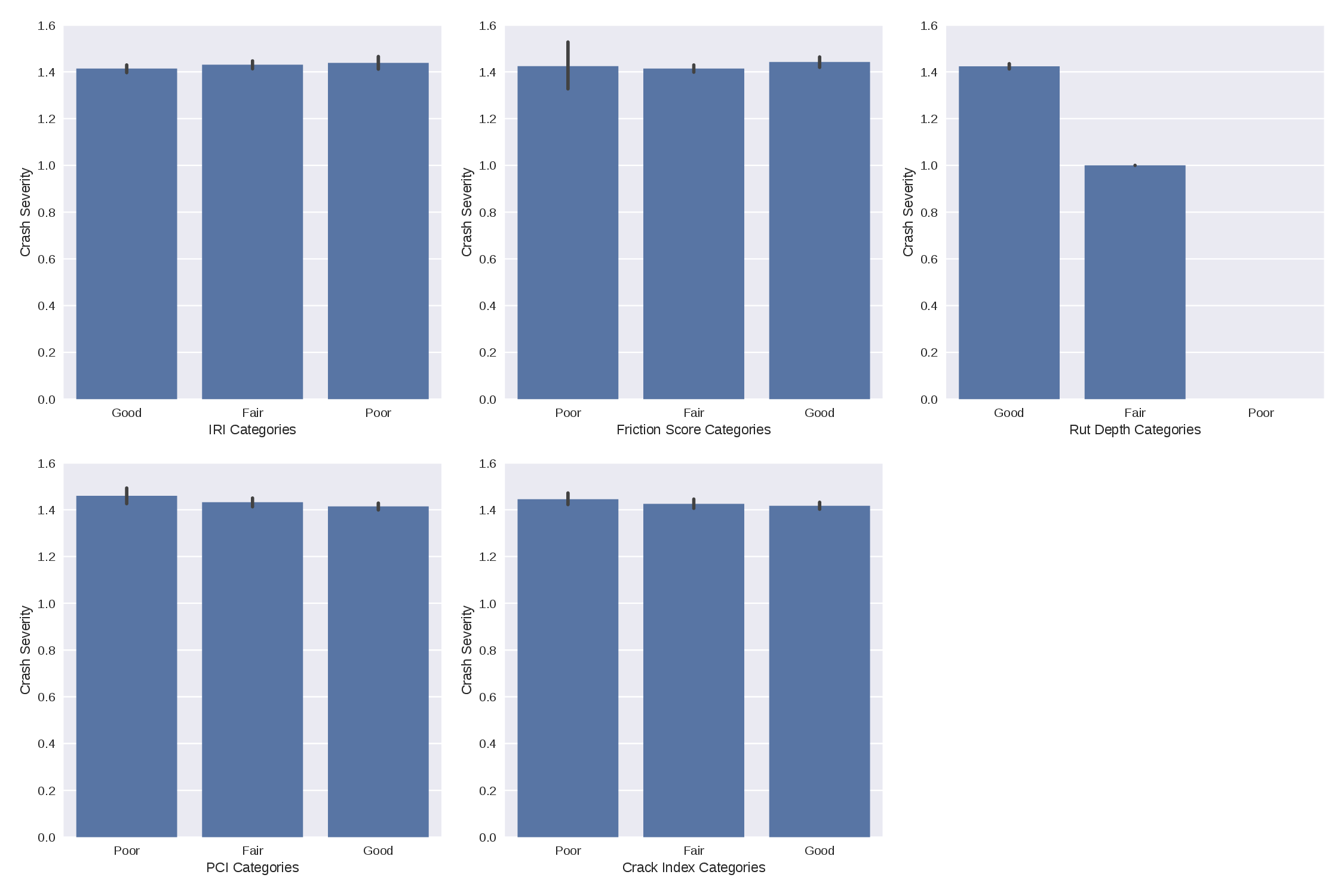}
    \caption{Pavement Condition vs. Crash Severity}\label{fig_pavement_crash}
\end{figure}

The crash severity is defined on a scale from 1 to 5, indicating property damage only, possible injury, minor injury, major injury, and fatal respectively. Similar to Figure~\ref{fig_cr_pavement}, Figure~\ref{fig_pavement_crash} presents the relationship between various pavement condition metrics and crash severity. The analysis reveals that IRI, FRICT, PCI\_2, and CRACK\_INDEX have minimal impact on crash severity, as the distribution of average crash severity across these categories is relatively uniform. Conversely, the data for rutting depth (RUT) shows that shallower rutting depths are associated with higher crash severity. This is likely because significant rutting often occurs in areas where driving speeds are lower, and drivers tend to be more cautious, resulting in lower crash severity. Therefore, it would be beneficial to investigate the combined impact of speed and pavement conditions on crash rates and severity, as discussed in later sections of this paper.

\subsubsection{Crash Severity Analysis by IRI}\label{iri-1}

Table~\ref{tab_iri_cs} presents the results of mean comparisons using the Tukey test to examine the relationship between IRI and crash severity. It shows that overall crash severity is not strongly influenced by IRI. The comparison does not return significant results for almost all scenarios except for single vehicle collisions, clear weather conditions, and dry surface conditions. In single vehicle collisions, rougher pavement is associated with more severe crashes, likely because rough pavement can lead to a loss of vehicle control, especially when a driver is already at risk of an accident. In clear weather conditions and on dry surface roads, rougher pavement is associated with less severe crashes, possibly because drivers tend to be more cautious on visibly rough surfaces, reducing their speed and increasing attention, which helps mitigate the severity of crashes.

\begin{table}[H]
    \caption{Mean Crash Severities by IRI Groups}\label{tab_iri_cs}
    \footnotesize
    \centering
\setlength{\tabcolsep}{3pt}
\renewcommand{\arraystretch}{0.85}    
    \begin{tabular}{|l|l|ccc|ccc|}
        \hline
        Category & Scenario & Good & Fair & Poor & G-F & F-P & G-P \\
        \hline
        \rule{0pt}{10pt}Overall & Overall Crash Severity & 1.413 & 1.430 & 1.438 & 0.3315 & 0.9017 & 0.3156 \\
        \hline
        \multirow{8}{*}{Collision Type} 
        & \rule{0pt}{10pt}Single Vehicle & 1.43 & 1.50 & 1.65 & 0.0079 & 0.0071 & 0.0000 \\
        & \rule{0pt}{10pt}Head On & 2.57 & 2.10 & 1.95 & 0.0111 & 0.7075 & 0.0070 \\
        & \rule{0pt}{10pt}Rear End & 1.53 & 1.42 & 1.38 & 0.0000 & 0.3617 & 0.0000 \\
        & \rule{0pt}{10pt}Angle Left Turn & 1.62 & 1.52 & 1.46 & 0.3579 & 0.6275 & 0.1075 \\
        & \rule{0pt}{10pt}Broadside & 1.80 & 1.65 & 1.56 & 0.0012 & 0.0810 & 0.0000 \\
        & \rule{0pt}{10pt}Sideswipe (Same) & 1.25 & 1.22 & 1.16 & 0.3148 & 0.0933 & 0.0029 \\
        & \rule{0pt}{10pt}Rear to Rear & 1.63 & 1.23 & 1.30 & 0.4597 & 0.9782 & 0.6218 \\
        & \rule{0pt}{10pt}Rear to Side & 1.07 & 1.26 & 1.27 & 0.3494 & 0.9990 & 0.5535 \\
        \hline
        \multirow{7}{*}{Weather Condition} 
        & \rule{0pt}{10pt}Clear & 1.57 & 1.53 & 1.47 & 0.0362 & 0.0375 & 0.0000 \\
        & \rule{0pt}{10pt}Cloudy & 1.52 & 1.48 & 1.41 & 0.4528 & 0.2454 & 0.0336 \\
        & \rule{0pt}{10pt}Freezing Rain & 1.30 & 1.41 & 1.28 & 0.4917 & 0.7031 & 0.9862 \\
        & \rule{0pt}{10pt}Rainy & 1.46 & 1.47 & 1.40 & 0.9793 & 0.6071 & 0.6802 \\
        & \rule{0pt}{10pt}Snow & 1.30 & 1.32 & 1.25 & 0.9152 & 0.5855 & 0.7353 \\
        & \rule{0pt}{10pt}Fog/Smoke & 2.27 & 1.75 & 1.28 & 0.1265 & 0.6158 & 0.1010 \\
        & \rule{0pt}{10pt}Severe Winds & 1.32 & 1.44 & 1.60 & 0.6913 & 0.9247 & 0.7672 \\
        \hline
        \multirow{7}{*}{Road Surface} 
        & \rule{0pt}{10pt}Dry & 1.58 & 1.53 & 1.47 & 0.0226 & 0.0148 & 0.0000 \\
        & \rule{0pt}{10pt}Wet & 1.49 & 1.48 & 1.36 & 0.9718 & 0.0413 & 0.0233 \\
        & \rule{0pt}{10pt}Ice/Frost & 1.32 & 1.39 & 1.31 & 0.2321 & 0.6030 & 0.9791 \\
        & \rule{0pt}{10pt}Snow & 1.35 & 1.29 & 1.32 & 0.4387 & 0.9411 & 0.8655 \\
        & \rule{0pt}{10pt}Slush & 1.26 & 1.27 & 1.61 & 0.9992 & 0.1833 & 0.1400 \\
        & \rule{0pt}{10pt}Sand & 3.00 & 2.00 & 2.00 & 0.8855 & 1.0000 & 0.8553 \\
        & \rule{0pt}{10pt}Gravel & 1.52 & 1.62 & 1.50 & 0.9545 & 0.9805 & 0.9992 \\
        \hline
        \multirow{6}{*}{Location} 
        & \rule{0pt}{10pt}On Roadway & 1.49 & 1.47 & 1.44 & 0.5116 & 0.2589 & 0.0411 \\
        & \rule{0pt}{10pt}Shoulder & 1.53 & 1.66 & 1.64 & 0.1385 & 0.9900 & 0.8313 \\
        & \rule{0pt}{10pt}Median & 1.30 & 1.38 & 1.60 & 0.4787 & 0.8204 & 0.6714 \\
        & \rule{0pt}{10pt}Roadside & 1.63 & 1.53 & 1.36 & 0.6097 & 0.6888 & 0.3764 \\
        & \rule{0pt}{10pt}Parking Lane & 1.30 & 1.18 & 1.28 & 0.7182 & 0.8391 & 0.9943 \\
        & \rule{0pt}{10pt}Left Turn Lane & 1.00 & 1.47 & 1.00 & 0.6028 & 0.1951 & 1.0000 \\
        \hline
        \multirow{6}{*}{Road Factors} 
        & \rule{0pt}{10pt}Wet/Icy & 1.35 & 1.38 & 1.34 & 0.6159 & 0.7745 & 0.9900 \\
        & \rule{0pt}{10pt}Debris & 1.13 & 1.38 & 1.00 & 0.1459 & 0.2752 & 0.8336 \\
        & \rule{0pt}{10pt}Work Zone & 1.55 & 1.45 & 1.34 & 0.5722 & 0.7757 & 0.4141 \\
        & \rule{0pt}{10pt}Slippery/Worn & 1.50 & 1.30 & 1.15 & 0.4320 & 0.7757 & 0.2588 \\
        & \rule{0pt}{10pt}Ruts/Holes & 1.60 & 1.00 & 1.00 & 0.5349 & 1.0000 & 0.7447 \\
        & \rule{0pt}{10pt}Obstruction & 1.11 & 1.63 & 1.50 & 0.0991 & 0.9644 & 0.7253 \\
        \hline
        \multirow{7}{*}{Drug/Alcohol} 
        & \rule{0pt}{10pt}Only Drugs & 3.02 & 2.92 & 2.28 & 0.9651 & 0.5983 & 0.4781 \\
        & \rule{0pt}{10pt}Alcohol (< Stat.) & 2.22 & 2.76 & 1.00 & 0.5900 & 0.2910 & 0.5348 \\
        & \rule{0pt}{10pt}Alcohol (Stat.) & 1.81 & 1.82 & 1.60 & 0.9983 & 0.5161 & 0.5242 \\
        & \rule{0pt}{10pt}Under Influence & 2.51 & 2.35 & 1.76 & 0.7127 & 0.1618 & 0.0440 \\
        & \rule{0pt}{10pt}Drug/Alc (< Stat.) & 2.80 & 3.00 & 3.50 & 0.9682 & 0.8920 & 0.8015 \\
        & \rule{0pt}{10pt}Drug/Alc (Stat.) & 4.00 & 4.00 & 3.75 & 1.0000 & 0.9351 & 0.9105 \\
        & \rule{0pt}{10pt}Refused & 1.71 & 1.66 & 1.61 & 0.9566 & 0.9739 & 0.9055 \\
        \hline
        \multirow{6}{*}{Lighting} 
        & \rule{0pt}{10pt}Daylight & 1.52 & 1.49 & 1.42 & 0.2537 & 0.0025 & 0.0000 \\
        & \rule{0pt}{10pt}Dusk & 1.55 & 1.37 & 1.55 & 0.0694 & 0.3041 & 0.9980 \\
        & \rule{0pt}{10pt}Dark (Lighted) & 1.50 & 1.47 & 1.49 & 0.8189 & 0.9324 & 0.9801 \\
        & \rule{0pt}{10pt}Dark (Unlit) & 1.54 & 1.61 & 1.77 & 0.2445 & 0.4364 & 0.1682 \\
        & \rule{0pt}{10pt}Dawn & 1.56 & 1.46 & 1.29 & 0.5535 & 0.5523 & 0.2344 \\
        & \rule{0pt}{10pt}Dark (Unknown) & 1.48 & 1.66 & 2.00 & 0.6571 & 0.6415 & 0.3252 \\
        \hline
        \multirow{5}{*}{Environment} 
        & \rule{0pt}{10pt}Weather & 1.37 & 1.36 & 1.34 & 0.9464 & 0.8974 & 0.7964 \\
        & \rule{0pt}{10pt}Visual Obst. & 1.54 & 1.52 & 1.51 & 0.9774 & 0.9979 & 0.9684 \\
        & \rule{0pt}{10pt}Glare & 1.65 & 1.70 & 1.65 & 0.9649 & 0.9784 & 0.9996 \\
        & \rule{0pt}{10pt}Animal & 1.18 & 1.11 & 1.05 & 0.0874 & 0.8087 & 0.3700 \\
        & \rule{0pt}{10pt}None Apparent & 1.56 & 1.53 & 1.46 & 0.6571 & 0.6415 & 0.3252 \\
        \hline
    \end{tabular}
\end{table}

\subsubsection{Crash Severity Analysis by Friction}\label{friction-1}

Table~\ref{tab_cs_friction} presents the results of mean comparisons using the Tukey test to examine the relationship between friction scores and crash severity. Overall, crash severity shows minor differences across the friction categories, with mean values of 1.42 for ``good", 1.41 for ``fair", and 1.44 for ``poor". There are no scenarios where all comparison pairs show significant differences. However, in scenarios such as head-on collisions, clear weather conditions, dry road surface conditions, and daylight lighting conditions, the ``good" friction category shows a significant difference from the ``fair" and ``poor" categories. In these scenarios, pavement sections with ``good" friction are associated with more severe crashes than those with ``fair" and ``poor" friction. This could be because drivers may feel more confident and travel at higher speeds on roads with better friction, leading to more severe crashes when they occur. Additionally, in optimal conditions (clear weather, dry roads, daylight), other factors like speed and driver behavior might play a more critical role, overshadowing the benefits of good pavement friction.

\begin{table}[H]
    \caption{Mean Crash Severities by Friction Score Groups}\label{tab_cs_friction}
    \footnotesize
    \centering
    \begin{tabular}{|l|l|ccc|ccc|}
        \hline
        Category & Scenario & Good & Fair & Poor & G-F & F-P & G-P \\
        \hline
        \rule{0pt}{10pt}Overall & Overall Crash Severity & 1.42 & 1.41 & 1.44 & 0.0863 & 0.9848 & 0.9605 \\
        \hline
        \multirow{7}{*}{Collision Type} 
        & \rule{0pt}{10pt}Single Vehicle & 1.48 & 1.41 & 1.33 & 0.0106 & 0.9591 & 0.8620 \\
        & \rule{0pt}{10pt}Head On & 2.83 & 2.29 & 1.28 & 0.0105 & 0.1308 & 0.0107 \\
        & \rule{0pt}{10pt}Rear End & 1.69 & 1.42 & 1.30 & 0.0000 & 0.5260 & 0.0010 \\
        & \rule{0pt}{10pt}Angle Left Turn & 1.66 & 1.59 & 1.92 & 0.7864 & 0.3754 & 0.5722 \\
        & \rule{0pt}{10pt}Broadside & 2.03 & 1.68 & 2.58 & 0.0000 & 0.7163 & 0.0031 \\
        & \rule{0pt}{10pt}Sideswipe (Same) & 1.30 & 1.22 & 1.13 & 0.0610 & 0.8069 & 0.5218 \\
        & \rule{0pt}{10pt}Rear to Rear & 1.00 & 1.52 & 1.00 & 0.8468 & 0.7284 & 1.0000 \\
        \hline
        \multirow{6}{*}{Weather Condition} 
        & \rule{0pt}{10pt}Clear & 1.73 & 1.50 & 1.46 & 0.0000 & 0.8900 & 0.0048 \\
        & \rule{0pt}{10pt}Cloudy & 1.64 & 1.45 & 1.33 & 0.0965 & 0.6679 & 0.0000 \\
        & \rule{0pt}{10pt}Freezing Rain & 1.41 & 1.28 & 1.50 & 0.3687 & 0.9126 & 0.9861 \\
        & \rule{0pt}{10pt}Rainy & 1.63 & 1.41 & 1.80 & 0.0059 & 0.5245 & 0.8886 \\
        & \rule{0pt}{10pt}Snow & 1.42 & 1.27 & 1.00 & 0.0043 & 0.5796 & 0.2744 \\
        & \rule{0pt}{10pt}Blowing Snow & 1.40 & 1.34 & 1.00 & 0.7750 & 0.8709 & 0.8339 \\
        \hline
        \multirow{4}{*}{Road Surface} 
        & \rule{0pt}{10pt}Dry & 1.73 & 1.50 & 1.39 & 0.0000 & 0.3336 & 0.0001 \\
        & \rule{0pt}{10pt}Wet & 1.66 & 1.43 & 1.75 & 0.0000 & 0.1981 & 0.8865 \\
        & \rule{0pt}{10pt}Ice/Frost & 1.45 & 1.30 & 1.20 & 0.0022 & 0.9463 & 0.7063 \\
        & \rule{0pt}{10pt}Snow & 1.42 & 1.30 & 1.28 & 0.0586 & 0.9963 & 0.8715 \\
        \hline
        \multirow{4}{*}{Location} 
        & \rule{0pt}{10pt}On Roadway & 1.57 & 1.45 & 1.44 & 0.0000 & 0.9859 & 0.1251 \\
        & \rule{0pt}{10pt}Shoulder & 1.82 & 1.44 & 1.00 & 0.0000 & 0.7808 & 0.4295 \\
        & \rule{0pt}{10pt}Off Traffic & 1.76 & 1.76 & 1.00 & 0.9980 & 0.4449 & 0.4406 \\
        & \rule{0pt}{10pt}Parking Lane & 1.22 & 1.26 & 1.00 & 0.9719 & 0.8602 & 0.9062 \\
        \hline
        \multirow{2}{*}{Road Factors} 
        & \rule{0pt}{10pt}Wet/Icy & 1.46 & 1.32 & 1.11 & 0.0000 & 0.6497 & 0.3008 \\
        & \rule{0pt}{10pt}Slippery/Worn & 1.55 & 1.23 & 1.00 & 0.1348 & 0.9404 & 0.7151 \\
        \hline
        \multirow{3}{*}{Drug/Alcohol} 
        & \rule{0pt}{10pt}Only Drugs & 2.89 & 3.00 & 1.00 & 0.8812 & 0.3913 & 0.4606 \\
        & \rule{0pt}{10pt}Alcohol (Stat.) & 1.89 & 1.77 & 1.00 & 0.6507 & 0.6148 & 0.5200 \\
        & \rule{0pt}{10pt}Refused & 1.90 & 1.63 & 1.00 & 0.4112 & 0.7886 & 0.6206 \\
        \hline
        \multirow{5}{*}{Lighting} 
        & \rule{0pt}{10pt}Daylight & 1.70 & 1.44 & 1.41 & 0.0000 & 0.9095 & 0.0004 \\
        & \rule{0pt}{10pt}Dusk & 1.55 & 1.50 & 1.50 & 0.8650 & 1.0000 & 0.9927 \\
        & \rule{0pt}{10pt}Dark (Lighted) & 1.66 & 1.44 & 1.55 & 0.0006 & 0.8358 & 0.8216 \\
        & \rule{0pt}{10pt}Dark (Unlit) & 1.61 & 1.55 & 1.00 & 0.3792 & 0.8409 & 0.8094 \\
        & \rule{0pt}{10pt}Dawn & 1.66 & 1.47 & 1.00 & 0.1395 & 0.6491 & 0.4322 \\
        \hline
        \multirow{3}{*}{Environment} 
        & \rule{0pt}{10pt}Weather & 1.50 & 1.30 & 1.16 & 0.0000 & 0.8869 & 0.4919 \\
        & \rule{0pt}{10pt}Glare & 2.52 & 1.35 & 1.00 & 0.0000 & 0.8669 & 0.0816 \\
        & \rule{0pt}{10pt}None Apparent & 1.72 & 1.50 & 1.44 & 0.0000 & 0.6889 & 0.0005 \\
        \hline
    \end{tabular}
\end{table}

\subsection{Impact of Speed Limits on Crash Rate and Severity}\label{integrating-speed-limits}

\begin{figure}[H]
    \centering
    \includegraphics[width=0.8\textwidth]{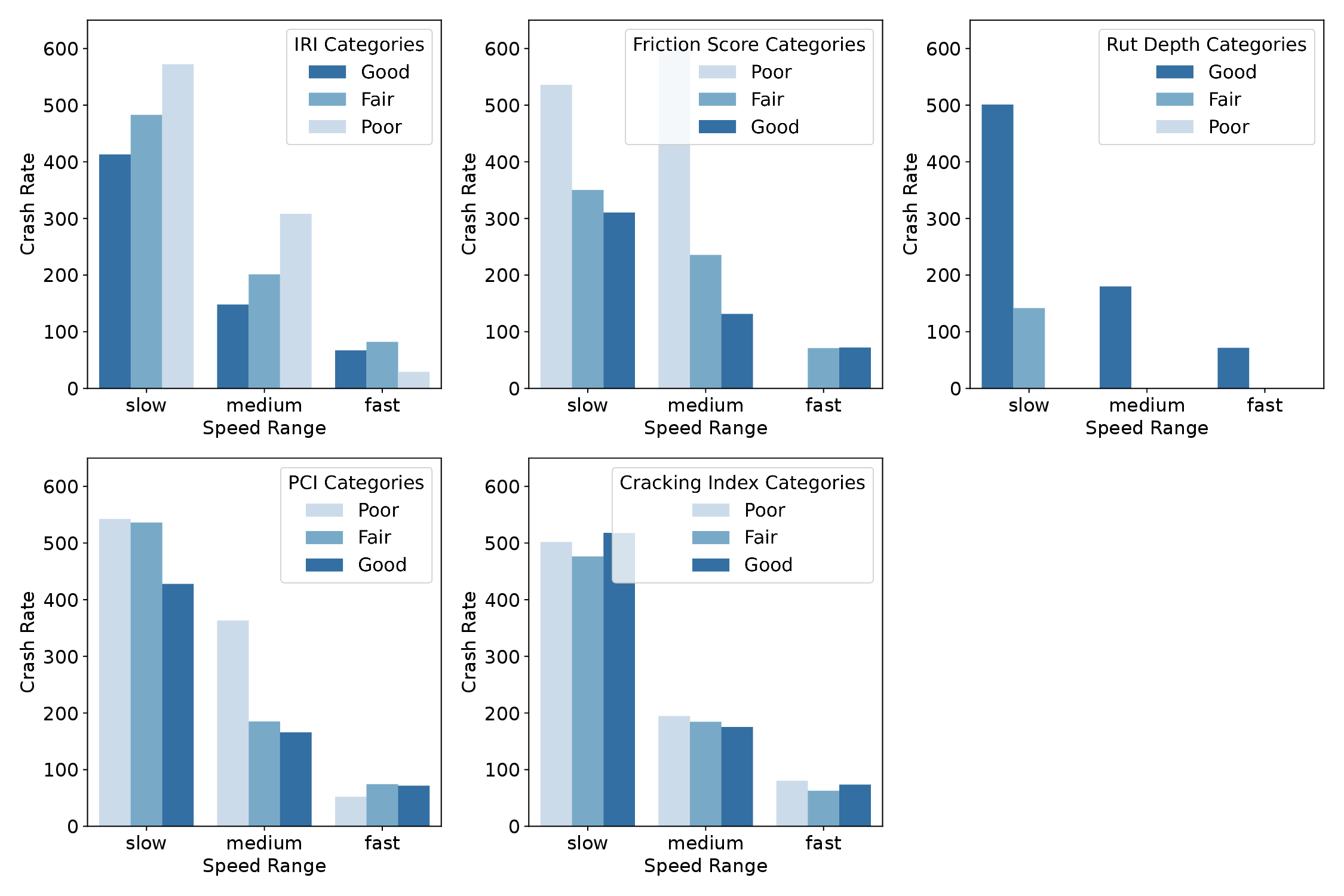}
    \caption{Mean crash rate for different pavement condition values with different speed limit}\label{fig_cr_vmt_iri}
\end{figure}

We also compare the mean crash rates and crash severity against pavement condition metrics in different speed limit ranges. Based on the distribution of the speed limit histogram in Figure \ref{fig_pavement_histogram}, roads with speed limits of 0-45 mph were classified as ``slow speed", 50-60 mph as ``medium speed", and 65 mph or higher as ``fast speed". Figure \ref{fig_cr_vmt_iri} depicts the relationship between various pavement condition metrics and crash rates across the previously defined speed ranges: slow, medium, and fast.

Based on this figure, we can observe that overall, pavement sections with higher speed limits are associated with lower crash rates. This is likely because roads with higher speed limits typically have better controlled access, superior design, consistent traffic flow, fewer intersections, advanced traffic control, and fewer distractions. The first subplot shows that for slow and medium speed ranges, smoother pavement is associated with lower crash rates. However, in the fast speed range, there is no clear trend between IRI and crash rates, with results even suggesting that rougher pavement may be associated with lower crash rates. For friction scores (FRICT), there is a similar trend where better friction is linked to lower crash rates in the slow and medium speed ranges, but no clear trend in the fast speed range. Regarding rutting depth (RUT), the data is insufficient for a comprehensive analysis, though it indicates that for low-speed ranges, more severe rutting depth correlates with lower crash rates. The pavement condition index (PCI\_2) shows a trend similar to IRI, where better pavement conditions are associated with lower crash rates in slow and medium speed ranges. For the cracking index (CRACK\_INDX), there is no clear trend between crash rates and cracking index categories.

% Figure~\ref{fig_cr_vmt_skid} illustrates the relationship between various pavement condition metrics and crash severity, segmented by speed ranges. Across different pavement metric ranges, no clear trend emerges except for friction (FRICT)  values within the slow speed limit range. Notably, a trend is observed wherein pavements with higher skid resistance in the slow speed range correspond to higher crash severity. 

\begin{figure}[H]
    \centering
    \includegraphics[width=0.8\textwidth]{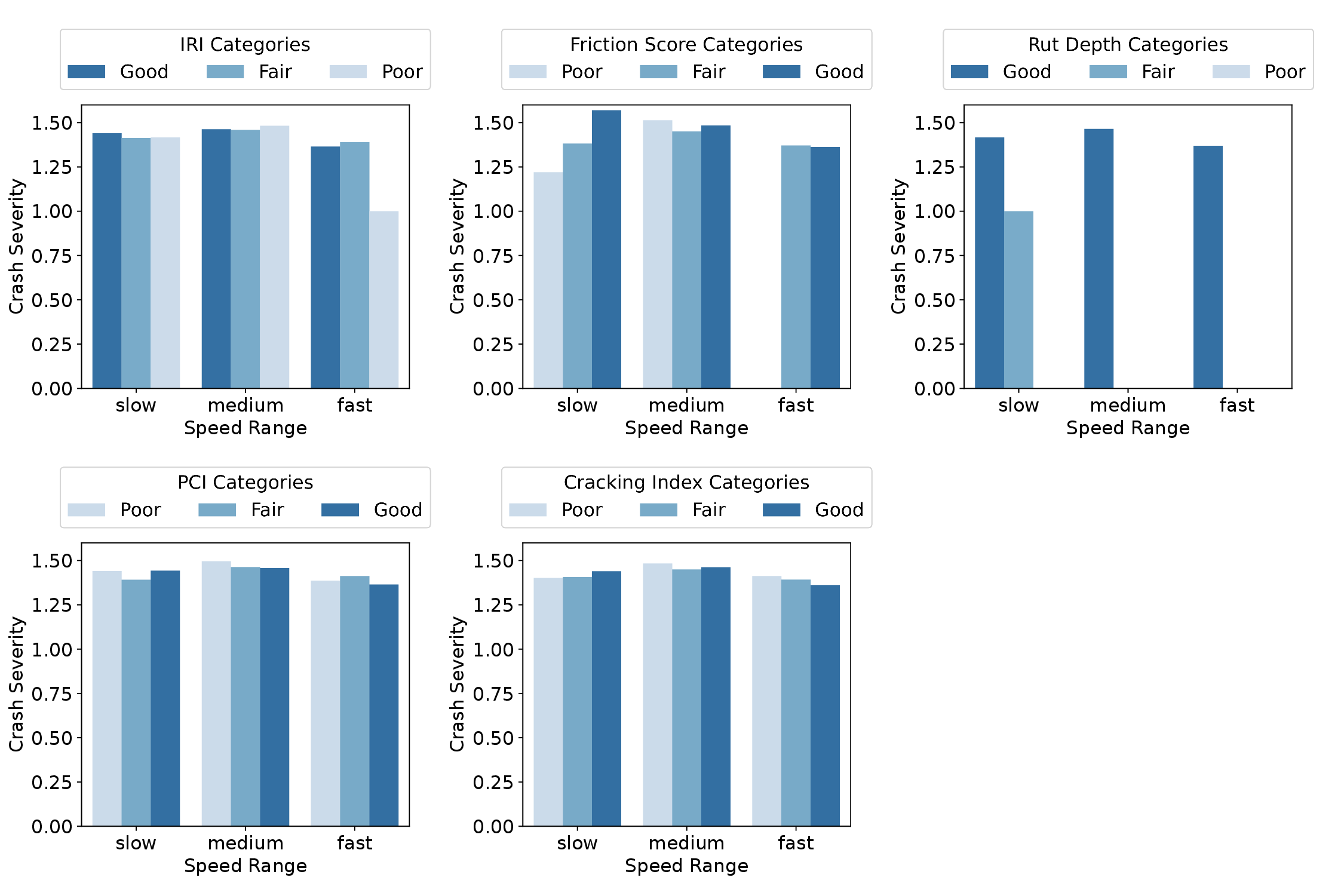}
    \caption{Mean crash severity for different pavement condition values with different speed limit}\label{fig_cs_vmt_skid}
\end{figure}

Figure~\ref{fig_cs_vmt_skid} illustrates the relationship between various pavement condition metrics and crash severity, segmented by speed ranges. Overall, no clear trend emerges across different pavement metric ranges, except for friction (FRICT) values within the slow speed limit range. Notably, a trend is observed wherein pavements with higher skid resistance in the slow speed range correspond to higher crash severity. This unexpected result may be due to several factors: drivers might feel overconfident on higher-friction surfaces at lower speeds, leading to riskier driving behaviors; or higher-friction surfaces might be more commonly found in areas with other contributing risk factors such as dense traffic, intersections, and pedestrian activity.

\section{Regression and Classification Modeling}\label{regression-and-classification-modeling}

\subsection{Crash Rate}\label{crash-rate-1}

\begin{figure}[H]
    \centering
    \includegraphics[width=0.7\textwidth]{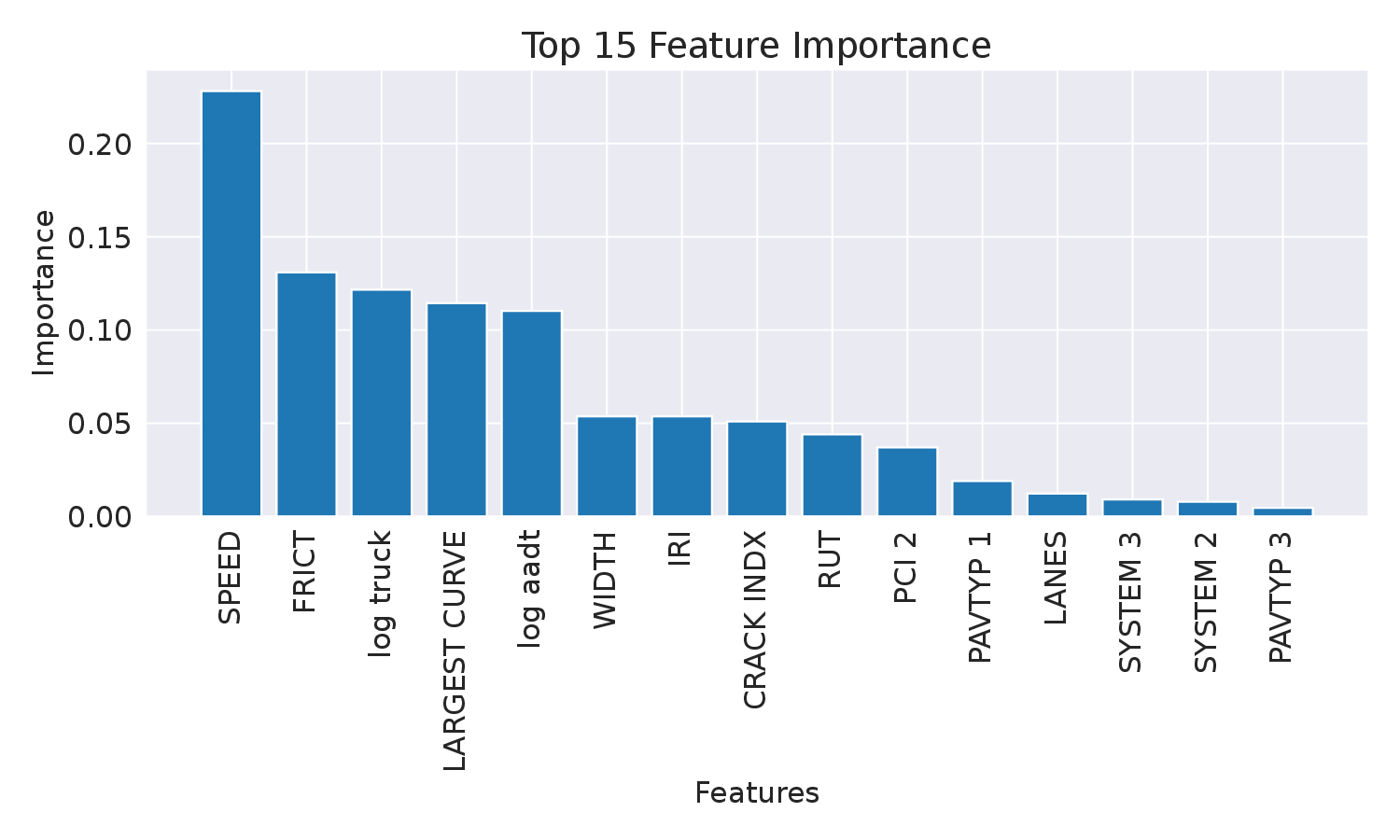}
    \caption{Top Features on Crash Rate}\label{fig_cr_features}
\end{figure}

We use random forest regression model to analyze the dataset, and the top features and their partial dependence were identified. Figure~\ref{fig_cr_features} shows the top 15 feature importances identified by a random forest regression model analyzing the dataset with crash rate as the target variable. The most important feature is SPEED, suggesting that the speed limit of a road segment heavily influences crash rates. Other notable features include FRICT, log\_truck, LARGE\_CURVE, and log\_aadt. The WIDTH, RUT (rut depth), IRI, CRACK\_RATIO (cracking ratio) and PCI\_2 (Pavement Condition Index) also have significant impact. Other features like PAVTYP\_1, LANES, SYSTEM\_3, SYSTEM\_2, and PAVTYP\_3 show that road design and functional class also contribute to crash rates.

\begin{figure}[H]
    \centering
    \includegraphics[width=\textwidth]{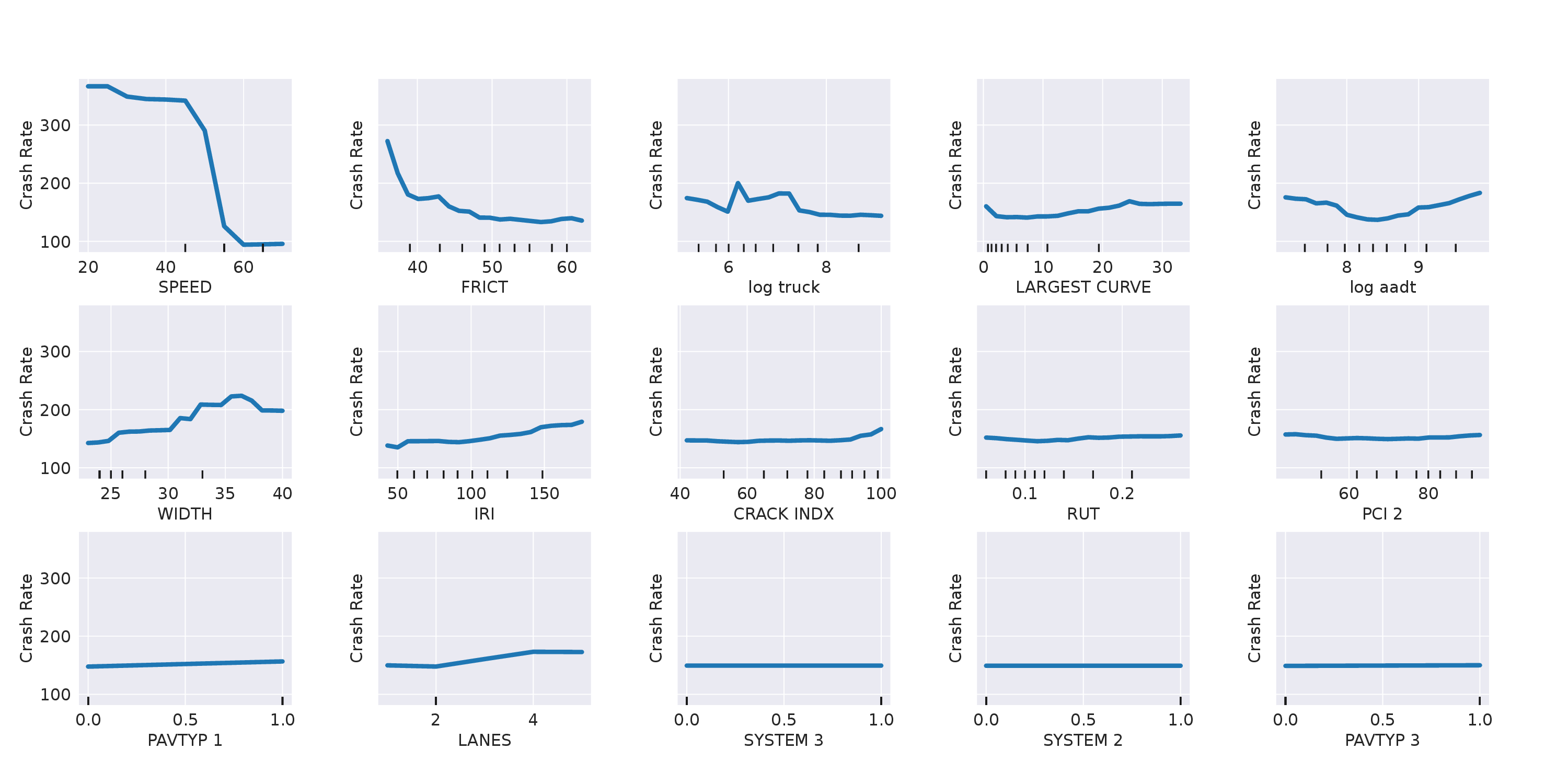}
    \caption{Partial dependence of top features on crash rate}\label{fig_cr_partial_dependence}
\end{figure}

Figure~\ref{fig_cr_partial_dependence} presents partial dependence plots for the top features identified by the random forest regression model analyzing crash rates. A Partial Dependence Plot visually represents the relationship between the features and the predicted target of a machine learning model, showing how the predicted probability or response variable changes with varying feature values. It helps to isolate the effect of specific features, making it easier to interpret and understand their influence on the model's predictions. The SPEED plot shows a steep decline in crash rates as speed increases from 35 to 65 mph. This decline can be attributed to the fact that higher speed limits are typically associated with well-designed roads, suggesting a correlation between higher speed limits on better-designed roads and lower crash rates. The FRICT plot reveals that crash rates decrease with increasing friction values, particularly between 30 and 50. The log\_truck and log\_aadt plot do not indicate obvious trend. The LARGEST\_CURVE plot shows slight increase in crash rate as the curve degree increases. The WIDTH plot reveals a slight increase in crash rates with road width, potentially due to drivers are more cautious in narrower road. The IRI plot indicates that crash rates increase with higher IRI values. The CRACK\_INDX, RUT, PCI\_2 and the rest of the features plot shows minimal variation, indicating a lesser impact on crash rates.

We also apply the Negative Binomial Regression (NBR) to the dataset. 
% NBR is an advancement of the Poisson regression model to address the overdispersion issue in count data modeling \cite{hilbe2014modeling}. The NBR model assumes that
% \begin{equation}
%     \log(\mu) = \beta_0 + \beta_1 x_1 + \beta_2 x_2 + \ldots + \beta_p x_p + \varepsilon
% \end{equation}\\
% \noindent Where,\\
% \(\mu\) = the expected value of the response variable (count data);\\
% \(\beta_0, \beta_1, \ldots, \beta_p\) = the regression coefficients to be estimated;  \\
% \(x_1, x_2, \ldots, x_p\) = the predictor variables.\\
% $\varepsilon$ = an error term following gamma distribution.
% This model form is suitable for crash prediction as it satisfies the criteria of predicting non-negative accident frequencies and logically predicts zero accidents when there is zero exposure. 
% Negative binomial regression is used when the data exhibit overdispersion, meaning the variance is greater than the mean, which is often the case in crash data \cite{lu2016accident}. 
The selection of variables in NBR is guided by the Akaike Information Criterion (AIC). Table~\ref{tab_regression} presents the results of a negative binomial regression model analyzing crash rates under four different scenarios: overall data, slow speed roads, medium speed roads, and fast speed roads. The coefficients, their estimates, and p-values are provided for each scenario. From Table~\ref{tab_regression}, in the overall negative binomial regression, the estimated coefficient on SPEED is –0.0175 (\(p < 0.001\)), indicating that—after adjusting for AADT (log\_aadt), heavy‐vehicle share (log\_truck), pavement friction (FRICT), geometric curvature (LARGEST\_CURVE), roughness (IRI), lane width (WIDTH), and other controls—each 1 mph increase in posted limit corresponds to a 1.7 percent reduction in expected crash frequency. This finding is also supported by the partial‐dependence plot in Figure \ref{fig_cr_partial_dependence}a, where predicted crash rates decline from roughly 300 to 100 crashes per 100 million vehicle-miles traveled as speed limits increase from 40 to 60 mph, demonstrating a robust negative trend across the random forest ensemble. The most plausible explanation is that higher‐speed facilities (e.g., controlled-access freeways and major arterials) are built and maintained to better geometric and safety standards, which outweigh the inherent risk of higher speeds. The variable log\_truck has a significant negative impact on crash rate in medium and fast speed limits but not in slow speed limits. This is likely due to other drivers being more aware and cautious around trucks on higher speed roads, maintaining greater distances and practicing safer driving behaviors, thus reducing the likelihood of crashes. The variable log\_aadt has a significant positive impact in all scenarios, likely reflecting that higher average annual daily traffic indicates more vehicles on the road, increasing the likelihood of crashes. Friction (FRICT) has a significant negative impact on crash rates in all scenarios, suggesting that better friction reduces the likelihood of vehicles skidding or losing control, thus lowering crash rates. The largest horizontal curve degree (LARGEST\_CURVE) has a significant positive impact in medium and fast speed limits but not in the slow speed limit, likely due to sharper curves at higher speeds increasing the risk of crashes due to greater vehicle instability. IRI has a positive impact in the medium speed range but not in the slow and fast speed ranges, possibly because medium speed roads are more sensitive to roughness, affecting vehicle control and increasing crash rates. This may also be because, on slow speed roads, the lower speeds allow drivers to better navigate rough surfaces, reducing the impact of roughness on crash rates. On fast speed roads, the overall design and maintenance tend to be better, minimizing the effects of rough surfaces on vehicle control and crash rates. Pavement width (WIDTH) has a significant positive impact in the medium and fast speed ranges but not in the slow speed range, likely because wider roads at higher speeds might encourage higher speeds and more frequent lane changes, increasing crash risks. On the other hand, the number of lanes (LANES) has a significant negative impact in the medium speed range, which could be because more lanes distribute traffic better, reducing congestion and potential collisions. This apparent contradiction might be explained by the fact that while wider roads provide more space and potentially encourage risky behavior, having more lanes can enhance traffic flow and reduce bottlenecks, thereby lowering the likelihood of crashes in the medium speed range.

\begin{table}[H]
    \caption{Negative Binomial Regression Results}
    \centering
    \label{tab_regression}
    \scriptsize
    \begin{tabular}{lllllllll}
        \hline
        \multicolumn{1}{c}{}              &
        \multicolumn{2}{c}{Overall}       &
        \multicolumn{2}{c}{Speed: Medium} &
        \multicolumn{2}{c}{Speed: Slow}   &
        \multicolumn{2}{c}{Speed: Fast}                                                                           \\
        \multicolumn{1}{c}{Coefficient}   &
        \multicolumn{1}{c}{Estimate}      &
        \multicolumn{1}{c}{p-value}       &
        \multicolumn{1}{c}{Estimate}      &
        \multicolumn{1}{c}{p-value}       &
        \multicolumn{1}{c}{Estimate}      &
        \multicolumn{1}{c}{p-value}       &
        \multicolumn{1}{c}{Estimate}      &
        \multicolumn{1}{c}{p-value}                                                                               \\\hline
        Intercept                         & -3.5356 & 0.000 & -3.9447 & 0.000 & -8.2101 & 0.000 & -4.3810 & 0.000 \\
        SYSTEM\_3{[}T.True{]}             & 0.0171  & 0.615 & 0.0334  & 0.401 & -0.0624 & 0.581 & 0.0632  & 0.493 \\
        PAVTYP\_1{[}T.True{]}             & 0.0562  & 0.078 & 0.0508  & 0.266 & -0.0450 & 0.702 & -0.0403 & 0.430 \\
        % PMIS\_LENGTH  & 0.1793  & 0.000 & 0.1510  & 0.000 & 0.5917  & 0.000 & 0.2256  & 0.000 \\
        SPEED                             & -0.0175 & 0.000 & N.A.    & N.A.  & N.A.    & N.A.  & N.A.    &       \\
        log\_truck                        & -0.1632 & 0.000 & -0.1878 & 0.000 & -0.0123 & 0.914 & -0.1081 & 0.045 \\
        log\_aadt                         & 1.0140  & 0.000 & 1.0069  & 0.000 & 1.2766  & 0.000 & 0.8312  & 0.000 \\
        FRICT                             & -0.0152 & 0.000 & -0.0190 & 0.000 & -0.0266 & 0.000 & -0.0111 & 0.004 \\
        LARGEST\_CURVE                    & 0.1380  & 0.000 & 0.1420  & 0.000 & 0.0475  & 0.538 & 0.1717  & 0.004 \\
        IRI                               & 0.0009  & 0.024 & 0.0015  & 0.002 & 0.0007  & 0.503 & 0.0018  & 0.060 \\
        WIDTH                             & 0.0053  & 0.144 & 0.0106  & 0.018 & -0.0039 & 0.647 & 0.0310  & 0.002 \\
        LANES                             & -0.0709 & 0.075 & -0.1455 & 0.008 & 0.0446  & 0.622 & -0.0077 & 0.932 \\
        alpha                             & 0.2826  & 0.000 & 0.3034  & 0.000 & 0.2019  & 0.000 & 0.2113  & 0.000 \\ \hline
    \end{tabular}
\end{table}

\subsection{Crash Severity}\label{crash-severity-2}

\begin{figure}[H]
    \centering
    \includegraphics[width=0.7\textwidth]{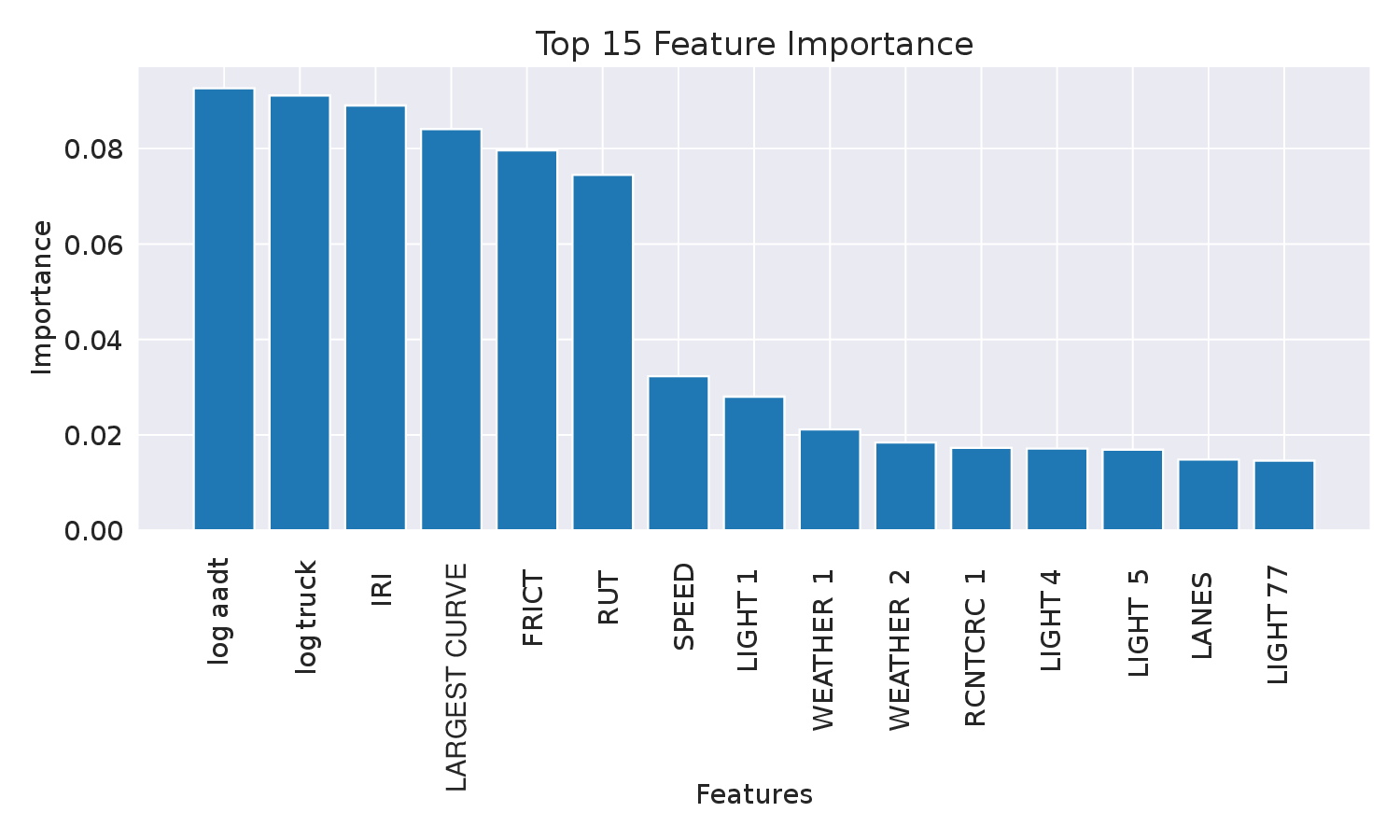}
    \caption{Top features on crash severity}\label{fig_cs_feasures}
\end{figure}

We applied the random forest classification model to the dataset to study the impact on crash severity models. Figure~\ref{fig_cs_feasures} presents the top 15 feature importances identified by a random forest classification model analyzing the impact on crash severity. The most important features include log\_aadt, log\_truck, IRI, FRICT, LARGEST\_CRUVE, FRICT, and RUT.

\begin{figure}[H]
    \centering
    \includegraphics[width=\textwidth]{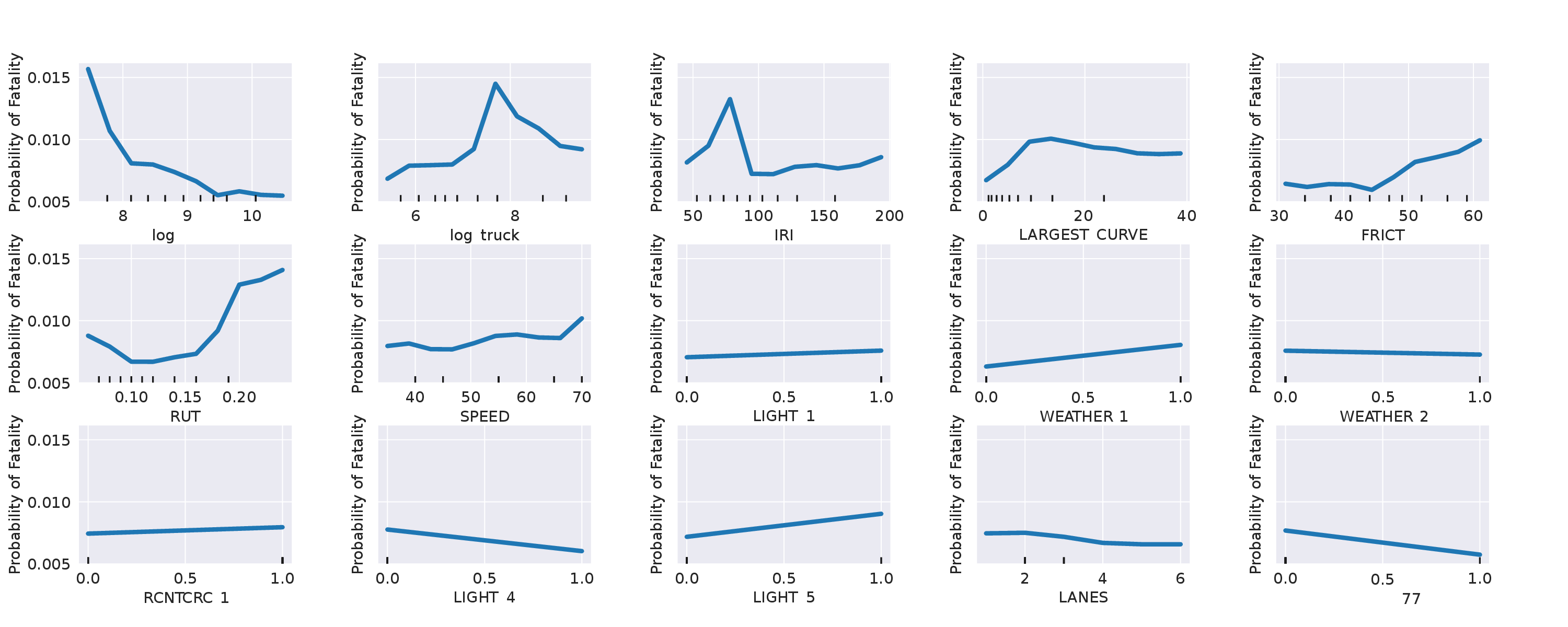}
    \caption{Partial dependence of top features on crash severity}\label{fig_cs_dependence}
\end{figure}

Figure~\ref{fig_cs_dependence} presents partial dependence plots for the top features' impact on the fatality class (the last class of the target variable). The y-axis represents the likelihood of being in the fatality class given the feature value. The plot for log\_aadt (log value of average annual daily traffic) shows that crash severity decreases with higher traffic volumes. The plots for log\_truck (log value of truck traffic) and IRI do not exhibit a clear trend. The LARGEST\_CURVE plot indicates a sharp increase in fatality likelihood when the largest horizontal curve degree increases from 0 to close to 15; thereafter, the fatality likelihood slightly decreases as the curve degree continues to increase. The FRICT (friction) plot shows that the fatality likelihood remains constant before a friction value of 45 and then increases as the friction value rises. The RUT plot displays a slight U-shape where the rutting depth around 0.13 corresponds to the lowest fatality likelihood. The SPEED plot reveals a generally increasing trend in crash severity with higher speed limits, suggesting that higher speed limits are associated with greater crash severity due to higher impact forces. The LIGHT\_5 (lighting condition being the 5th option: dark - roadway not lighted) plot shows that the likelihood of fatality is higher in dark scenarios than in other scenarios. Interestingly, the fatality likelihood is lower in the scenario of dark - roadway lighted. The WEATHER\_1 (weather condition being the first option: clear) plot shows that clear weather conditions are associated with a higher likelihood of fatality compared to other weather conditions.

% These insights highlight the importance of considering multiple factors in road design and traffic management to enhance road safety and mitigate crash severity.

We also used the ordered probit model to examine crash severity. The model segments the underlying severity variable into five distinct categories using five cut-off points. The model estimates a continuous hidden variable based on relevant factors. This hidden variable is then divided into sections using estimated cut-off points corresponding to each severity level. Table~\ref{tab_probit} presents the results of an ordered probit model analyzing crash severity across four scenarios: the overall dataset, medium speed limit roads, fast speed limit roads, and slow speed limit roads. Each row represents a variable's coefficient estimate and p-value, showing its impact on crash severity. The variables are selected based on the AIC to ensure the best model fit. The results show that higher friction values (FRICT) are associated with more severe crashes in the medium speed range, possibly because drivers feel more confident and might drive more aggressively on high-friction surfaces. Higher speeds (SPEED) increase crash severity due to greater impact forces during collisions. Crashes during daylight (LIGHT\_1) are associated with more severe outcomes, perhaps because higher traffic volumes and speeds during the day increase the risk and severity of crashes. Cloudy weather (WEATHER\_2) may contribute to more severe crashes, possibly due to reduced visibility and changing driving conditions that can lead to higher impact collisions. Clear weather conditions (WEATHER\_1) are linked to more severe crashes, which might be due to higher speeds and traffic volumes when weather conditions are favorable, increasing the likelihood of severe accidents. Dry road surfaces (CSRFCND\_1) see more severe crashes, possibly because drivers may drive faster on dry roads. Dark roads not lighted (LIGHT\_5) might lead to severe crashes due to poor visibility. In the ordered probit model (Table \ref{tab_probit}), the coefficient on log\_aadt is –0.0869 (\(p = 0.001\)), indicating that the increase in the natural log of daily traffic volume lowers the probability of a fatality-level crash. The partial‐dependence plot for log aadt (Figure \ref{fig_cs_dependence}a) further highlights this effect: as log\_aadt increases from approximately 7 to 10, the model’s predicted probability of a fatal crash declines from about 1.5 percent to 0.5 percent, demonstrating a clear downward trend. A possible explanation is that high‐volume facilities—such as interstates and major arterials—are designed with multiple lanes, median barriers, superior lighting, and more maintenance, all of which reduce the likelihood of fatal collisions. The ``not reported" category (ECNTCRC\_77) shows a similar trend, possibly due to data reporting issues or unmeasured factors.

It is interesting to note that FRICT (friction) is significant in the medium speed range but not in the slow and fast speed ranges. This may be because medium speed roads often have more variable conditions and traffic patterns, leading to a more pronounced effect of friction on crash severity. In contrast, on slow speed roads, the lower speeds reduce the overall impact of friction, while on fast speed roads, the friction is usually better maintained, and other factors such as high-speed maneuvers and driver reaction times may play a larger role in determining crash severity. It is also worth noting that IRI is not significant in any of the scenarios. This could be because the overall roughness of the pavement does not play a major role in crash severity compared to other factors such as speed, friction, and traffic conditions. RUT (rutting depth) is not significant in the overall, medium speed, and slow speed scenarios but is significant with a positive sign in the fast speed range. This may be because at higher speeds, deeper ruts can cause greater vehicle instability and more severe crashes, whereas at lower speeds, drivers can better navigate or avoid these road hazards.

\begin{table}[H]
    \caption{Ordered Probit Model Results}\label{tab_probit}
    \centering
    \scriptsize
    \begin{tabular}{lllllllll}
        \hline
        \multicolumn{1}{c}{}             &
        % \multicolumn{1}{c}{} &
        \multicolumn{2}{c}{Overall}      &
        \multicolumn{2}{c}{Medium Speed} &
        \multicolumn{2}{c}{Fast Speed}   &
        \multicolumn{2}{c}{Slow Speed}                                                                                 \\
        \multicolumn{1}{c}{Variable}     &
        \multicolumn{1}{c}{Coefficient}  &
        \multicolumn{1}{c}{p-value}      &
        \multicolumn{1}{c}{Coefficient}  &
        \multicolumn{1}{c}{p-value}      &
        \multicolumn{1}{c}{Coefficient}  &
        \multicolumn{1}{c}{p-value}      &
        \multicolumn{1}{c}{Coefficient}  &
        \multicolumn{1}{c}{p-value}                                                                                    \\ \hline
        log\_aadt                        & -0.0869    & 0.001 & -0.1151 & 0.001 & -0.1153 & 0.078 & 0.1265     & 0.114 \\
        largest\_curve                   & 0.0252     & 0.183 & 0.0058  & 0.795 & 0.0269  & 0.956 & 0.1156     & 0.118 \\
        IRI                              & -1.438e-05 & 0.960 & 0.0003  & 0.379 & 0.0011  & 0.142 & -2.801e-05 & 0.969 \\
        log\_truck                       & -0.0328    & 0.095 & 0.0258  & 0.328 & -0.0241 & 0.653 & -0.0997    & 0.186 \\
        FRICT                            & 0.0035     & 0.022 & 0.0043  & 0.036 & -0.0023 & 0.438 & 0.0049     & 0.277 \\
        RUT                              & 0.1477     & 0.425 & 0.1399  & 0.554 & 0.7502  & 0.041 & -0.4602    & 0.499 \\
        SPEED                            & 0.0065     & 0.000 & N.A.    & N.A.  & N.A.    & N.A.  & N.A.       & N.A.  \\
        LIGHT\_1                         & 0.1116     & 0.000 & 0.1714  & 0.000 & 0.0813  & 0.125 & -0.0100    & 0.909 \\
        WEATHER\_2                       & 0.2253     & 0.000 & 0.1906  & 0.002 & 0.2833  & 0.000 & -0.0936    & 0.551 \\
        WEATHER\_1                       & 0.2423     & 0.000 & 0.2307  & 0.000 & 0.2134  & 0.003 & 0.0440     & 0.785 \\
        % RDTYP\_1& 0.0777     & 0.014 & 0.0889  & 0.032 & 0.1075  & 0.064 & -0.0908    & 0.366 \\
        % CRCOMNNR\_3    & -0.2575    & 0.000 & -0.3736 & 0.000 & -0.0539 & 0.503 & -0.2167    & 0.046 \\
        % RDTYP\_12& 0.0229     & 0.516 & 0.0424  & 0.321 & 0.0553  & 0.630 & -0.0302    & 0.750 \\
        CSRFCND\_1                       & 0.1653     & 0.000 & 0.1445  & 0.004 & 0.2460  & 0.000 & 0.0700     & 0.603 \\
        LANES                            & 0.0295     & 0.114 & 0.0260  & 0.312 & 0.0832  & 0.091 & -0.0485    & 0.287 \\
        % CRCOMNNR\_5    & 0.0985     & 0.020 & 0.0126  & 0.802 & 0.5279  & 0.000 & 0.1056     & 0.342 \\
        ECNTCRC\_77                      & -2.2565    & 0.000 & -2.4028 & 0.000 & -5.1321 & 0.961 & -0.7618    & 0.013 \\
        LIGHT\_5                         & 0.2209     & 0.000 & 0.2653  & 0.000 & 0.1838  & 0.003 & 0.5572     & 0.013 \\
        % CRCOMNNR\_1    & -0.3928    & 0.000 & -0.3635 & 0.000 & -0.3869 & 0.000 & 0.0654     & 0.619 \\
        % CRCOMNNR\_6    & -0.7033    & 0.000 & -0.8479 & 0.000 & -0.5324 & 0.000 & -0.7353    & 0.000 \\
        1/2                              & 0.0230     & 0.938 & 0.5082  & 0.249 & -1.0188 & 0.276 & -0.3036    & 0.714 \\
        2/3                              & -0.5936    & 0.000 & -0.5474 & 0.000 & -0.7197 & 0.000 & -0.3676    & 0.000 \\
        3/4                              & -0.2134    & 0.000 & -0.2138 & 0.000 & -0.2181 & 0.000 & 0.0003     & 0.997 \\
        4/5                              & -0.5363    & 0.000 & -0.4966 & 0.000 & -0.5839 & 0.000 & -0.4776    & 0.060 \\ \hline
    \end{tabular}
\end{table}

\section{Conclusions}\label{conclusions}

This study provides a comprehensive analysis of the relationship between pavement conditions and traffic crash characteristics. By integrating crash data and pavement inventory data, we examined how various pavement related characteristics impact crash rates and severity. Statistical methods, including the Tukey test, negative binomial model, and ordered probit model, and machine learning models such as random forest were used in this study. 

Our analysis reveals that crash rates increase with higher IRI values, although smoother pavement is associated with lower crash rates in the slow and medium speed ranges, with no clear trend in the fast speed range. IRI has little impact on crash severity. Higher friction scores are linked to lower crash rates, especially in the slow and medium speed ranges, but not in high-speed areas. This is supported by the significant negative effect in both the NBR regression and random forest partial dependence plots. Friction significantly decreases crash rates particularly between values of 30 to 50. Based on the ordered probit model, friction also has a positive impact on crash severity in the medium speed range while not significant in the slow and fast speed ranges. Shallower rutting depths correlate with higher crash rates and severity, but in low-speed ranges, more severe rutting correlates with lower crash rates, with a depth of 0.13 inches corresponding to the lowest fatality likelihood. The Pavement Condition Index (PCI) shows a clear inverse relationship with crash rates, indicating better pavement condition corresponds to lower crash rates, though PCI minimally impacts severity. Similarly, a better cracking index is associated with fewer crashes, but the correlation is weak, and it minimally impacts severity.

The analysis also shows that higher speed limits consistently reduce crash rates.  However, the ordered probit model shows that higher speeds also increase crash severity. This highlights speed’s dual influence on safety outcomes. Moreover, traffic volume (log\_aadt) increases crash rates. Both random forest and NBR models show a positive relationship. However, higher traffic volume is linked to lower crash severity in the ordered probit model. This may be due to better infrastructure on busy roads. 

These findings suggest significant implications for road safety management and policy-making. The results highlight the importance of maintaining smooth pavement and adequate friction levels to reduce crash rates, particularly in slow and medium speed ranges. Addressing shallow rutting depths could help mitigate crash severity, especially on low-speed roads. Policymakers should prioritize pavement maintenance and improvement strategies that target these specific metrics to enhance overall road safety and reduce crash frequencies.

% \bibliography{sn-bibliography}% common bib file

\bibliographystyle{unsrtnat}
\bibliography{sn-bibliography} % Specify the name of your .bib file

\end{document}